\documentclass[aps, prb,
 amsmath,amssymb,
 twocolumn,
 floatfix,
reprint,
superscriptaddress,
longbibliography
]{revtex4-2}
\usepackage[english]{babel}
\usepackage{graphicx}
\usepackage{placeins} 
\usepackage{enumitem} 
\usepackage{dcolumn}
\usepackage{bm}
\usepackage{gensymb} 
\usepackage{xfrac}
\UseCollection{xfrac}{plainmath}
\usepackage{nicefrac}
\usepackage{cases}
\usepackage{upgreek}

\DeclareMathOperator{\sgn}{sgn}

\usepackage[normalem]{ulem}
\usepackage{xcolor}
\definecolor{eloicyan}{rgb}{0.,0.64,0.84}
\definecolor{ELOICYAN}{rgb}{0.,0.64,0.84}
\definecolor{darkgreen}{rgb}{0.,0.46,0.46}
\definecolor{lighgreen}{rgb}{0.2,0.75,0.56}
\definecolor{darkblue}{rgb}{0.1,0.2,0.46}
\definecolor{lightred}{rgb}{1.0,0.3,0.3}
\definecolor{darkred}{rgb}{.8,0.1,0.3}
\definecolor{lilac}{rgb}{0.6,0.2,0.7}

\renewcommand{\vec}[1]{\boldsymbol{\mathbf{#1}}}
\newcommand{\unitvec}[1]{\hat{\boldsymbol{\mathbf{#1}}}}

\DeclareMathAlphabet\mathbfcal{OMS}{cmsy}{b}{n} 
\DeclareUnicodeCharacter{2212}{-} 

\begin{document}


\title{Dynamics of bistable N\'{e}el domain walls under spin-orbit torque}

\author{Eloi Haltz}
\affiliation{School of Physics and Astronomy, University of Leeds, Leeds LS2 9JT, United Kingdom}
\affiliation{LSPM-CNRS, UPR3407, Sorbonne Paris Nord University, Villetaneuse, France}

\author{K\'{e}vin J. A. Franke}
\affiliation{School of Physics and Astronomy, University of Leeds, Leeds LS2 9JT, United Kingdom}

\author{Christopher H. Marrows}
\email{c.h.marrows@leeds.ac.uk}
\affiliation{School of Physics and Astronomy, University of Leeds, Leeds LS2 9JT, United Kingdom}

\date{\today}


\begin{abstract}
N\'{e}el magnetic domain walls that are stabilized by achiral energy terms instead of the usual Dzyaloshinskii-Moriya interaction will be bistable, with the two possible chiral forms being degenerate. Here we focus on the theoretical study of the spin-orbit torque driven dynamics of such bistable N\'{e}el domain walls. We find that, for a given domain wall, two propagation directions along a nanowire are possible, depending on its initial state. These dynamics also exhibit complex dependence on the spin-orbit torque magnitude, leading to important transient regimes. Finally, a few ways are proposed for controlled or random reversal of the domain wall propagation direction. A robust analytical model which handles all the observed behaviors of such domain walls is developed and validated by comparing with numerical simulations. The obtained new dynamics open the way for new uses of domain walls in information storage and processing devices.
\end{abstract}

\maketitle

\section{Introduction}

The electrical manipulation of magnetic domain walls (DWs) and their potential technological applications have motivated intense research in magnetism and spintronics in the last few decades~\cite{Kumar2022}. The possibility of encoding digital information in the magnetization and manipulating it by shifting the DWs along a nanowire opens new ways to construct data storage~\cite{Kumar2022,Parkin2008a}, logic devices~\cite{Allwood2005a,Lin2022} and, more recently, disruptive processing operations~\cite{Luo2019,Liu2021,Williame2021,Ababei2021,Siddiqui2020,Hassan2018}. However, these new uses of DWs require more and more complex dynamics. So far, different mechanism have been proposed for driving the DW motion in out-of-plane magnetized wires~\cite{Shibata2011}: applied magnetic fields~\cite{Beach2005} or electrical currents \textit{via} spin-transfer-torques (STTs)~\cite{Thiaville2005a} or, more efficiently, spin-orbit-torques (SOTs) for N\'{e}el DWs~\cite{Manchon2019}.

Driving the motion of a N\'{e}el DW under SOT in a ferromagnetic (FM)/heavy metal (HM) bilayer relies on the combination of two phenomena. First, the spin Hall effect (SHE) converts the electrical current going through the HM layer into a transverse pure-spin-current injected in the FM layer. The resulting anti-damping-like SOT drives N\'{e}el DWs in a direction which depends on their chirality \cite{Khvalkovskiy2013,Ryu2013,Thiaville2012}. The second is an effect that stabilizes these DWs in the N\'{e}el structure \cite{Chen2013,Torrejon2014,Hrabec2014}. Usually, that is achieved by the DMI emerging from the asymmetry between the interfaces of the FM layer which promotes the N\'{e}el type of the DWs with a given chirality, fixed by the FM/HM stack~\cite{Thiaville2012,Emori2013,Ryu2013}. In that classic scenario, the sign of the DMI sets the chirality of the N\'{e}el DW and thus fixes the propagation direction for a given SOT drive.

Other phenomena can also be used to stabilize N\'{e}el DWs, such as: an additional in-plane anisotropy \cite{Franke2021} or depth gradients of thickness~\cite{Wu2015}, strains~\cite{Chen2015,Dejong2015a,Belyaev2020,Fattouhi2022} or magnetization. Unlike the DMI, these effects stabilize N\'{e}el DWs but without any preferential chirality. One can then speak of bistable N\'{e}el DWs where the chiralities are energetically degenerate. As before, the DW dynamics depends on the internal structure of the DW so in this scenario, it can be reversed without changing the driving force if the chirality can be switched. This new paradigm opens up many possibilities, one being to process information directly encoded into the chirality of the DW itself. Binary operations based on this new degree of freedom have already been demonstrated numerically and experimentally for vortex DWs in in-plane magnetised Permalloy wires, driven by applied field \cite{Omari2019}. In that case, the wall chirality does not affect its field-driven dynamics. Here, new phenomena are described for out-of-plane magnetized systems under SOT, possibly field-free, where the direction of motion reverses for a given drive when the chirality is flipped. The  following results also demonstrate the additional use of these dynamics for random or quasi-random operations.

Here, the shape anisotropy resulting from the demagnetization effect in narrow magnetic wires is considered, allowing bistable N\'{e}el DWs~\cite{Skaugen2019,Boehm2017}. That effect is suitable for a multitude of well-known magnetic systems in thin films just by patterning structures, making it very convenient and versatile. To manipulate such a bistable N\'{e}el DW with SOT, the FM wire can be sandwiched between two HM layers but of different thicknesses: HM$_{\textrm{thick}}$/FM/HM$_{\textrm{thin}}$, as shown Fig.~\ref{fig:Fig1}(a). This cancels out any net DMI because of the symmetric FM interfaces and provides a non-zero net SHE as the electrical current density in each HM layers are different but with the same charge/spin conversion. Bistable N\'{e}el DWs have even already been imaged in such stack~\cite{Boehm2017}. These required properties can also be obtained by introducing a thin metallic spacer layer between the HM and the FM (such as Cu: HM/Cu$_{\textrm{thin}}$/FM) which prevents HM/FM direct interfaces but allows for spin injection though the thin metallic layer~\cite{Avci2019b,Nan2015}. More generally, the use of an HM with a significant SHE but without DMI is possible (such as Au, $\beta$-Ta or $\beta$-W).

\section{Methods}

In the following, the dynamics of these bistable N\'{e}el DWs under SOT is explored. A collective coordinate model is developed to describe the DW behaviors with two variables: $q$ - the DW position along the wire (and so its velocity $v=\dot{q}$) and $\varphi$ - the angle between the magnetization at the center of the DW and the current direction (with $\varphi=0$ or $\pi$ refering to a N\'{e}el DW and $\varphi=\pi/2$ or $-\pi/2$ corresponding to a Bloch DW) as shown Fig.~\ref{fig:Fig1}(a). For every following situation, the initial state of the DW is defined with two parameters: $Q=\pm1$ which corresponds to the orientation of the surrounding magnetic domains ($Q=1$ for an up/down DW and $Q=-1$ for a down/up DW), and $\varphi^0$ the initial orientation of the magnetization at the center of the DW. 

The obtained results are in very good agreement with numerical micromagnetic simulations that we have performed, validating some assumptions of the analytical model such as the DW shape or the demagnetization effects~\cite{Chen2002,Mougin2007,Skaugen2019,You2006b}. The micromagnetic simulations have been performed using the \textsc{Mumax3} open-source software allowing a finite difference approach to the magnetization dynamics \cite{Vansteenkiste2014}.  The magnetic wire is discretized into 2~nm $\times$ 2~nm $\times$ 1~nm cells. A saturation magnetization of $M_\textrm{s} = 1 \times 10^6$~A/m, an exchange stiffness of $A =3 \times 10^{-11}$~J/m, an out-of-plane uniaxial anisotropy of $K_0 = 1 \times10^6$~J/m$^3$, a gyromagnetic ratio of $\gamma_0 = 0.2 \times 10^6$~Hz/(A/m) and a Gilbert damping parameter of $\alpha=0.1$ have been considered \cite{Franke2021,Devolder2016}. The chosen a 32~nm width and a 1~nm thickness for the wire leads to an effective uniaxial shape anisotropy along the magnetic wire of magnitude $K_\textrm{IP} = 14$~kJ/m$^3$, which stabilizes bistable N\'{e}el DWs. The spin current $J_\textrm{s}$ is injected into the magnetic wire with a spin direction transvers to the wire ($\vec{\sigma}=\unitvec{y}$, as shown in Fig.~\ref{fig:Fig1}(a)). For simplicity, we assume a charge/spin conversion ratio of unity so that the charge and spin currents have equal magnitudes.

\section{Results}

\subsection{Domain wall dynamics under spin-orbit torque} 

\begin{figure}
    \includegraphics[width=8cm]{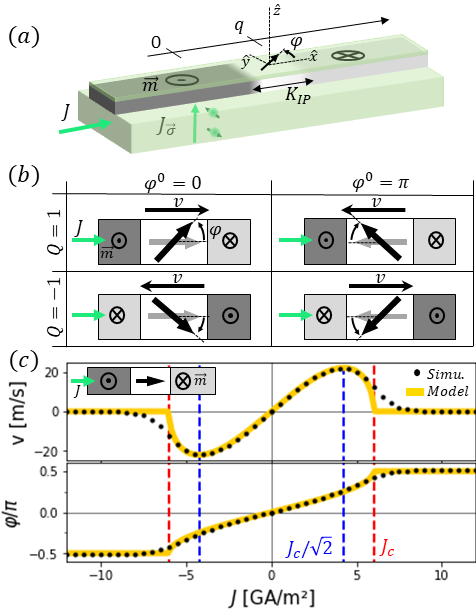}
    \caption{N\'{e}el wall motion under SOT. (a) Sketch of a bistable N\'{e}el DW in an asymmetric HM/FM/HM trilayer. (b) Sketch of the propagation direction and internal dynamics of a bistable N\'{e}el DW depending on the orientation of the surrounding domains (written as $Q=1$ for an up/down DW and $Q=-1$ for a down/up DW) and on the initial state of the DW ($\varphi^0=0$ or $\pi$, shown as a gray arrow). The tilting of the internal magnetization of the DW is shown as a black curved arrow. These four cases summarize the results obtained with micromagnetic simulations. (c) Evolution of the DW velocity $v$ (top panel) and associated internal angle $\varphi$ (bottom panel) with the current density $J$ for a DW with $Q=1$, $\varphi^0=0$. Black points correspond to the values extracted from the numerical simulations after 20~ns of current injection. The yellow full lines show the theoretical evolution of these quantities in the steady state regime $v^\infty$ and $\varphi^\infty$ given by Eq.~(\ref{eq:v_de_j_v}) and (\ref{eq:v_de_j_phi}), respectively. The red and the blue vertical dashed lines correspond to the critical currents $J_\textrm{c} = 6$~GA/m$^2$ and $\frac{J_\textrm{c}}{\sqrt{2}} = 4.25$~GA/m$^2$.
    \label{fig:Fig1}}
\end{figure}

First, the DW is set in a certain initial state ($Q$,$\varphi^0$) and a small spin current of $J = 3$~GA/m$^2$ is injected. The diagrams in Fig.~\ref{fig:Fig1}(b) show in which direction the DW moves and how $\varphi$ evolves under SOT depending on that ($Q$,$\varphi^0$) configuration. For a given $Q$ (rows of the table in Fig.~\ref{fig:Fig1}(b)), the DW moves in opposite directions depending on $\varphi_0$. Thus, just by reversing the initial orientation of the DW $\varphi^0$, its motion is reversed. Also, for a given chirality (diagonals of the table Fig.~\ref{fig:Fig1}(b)) the up/down and down/up DWs move in the same direction. That is the so-called coherent shift of DWs and it has previously been ensured by the DMI. Here, since the chirality is not fixed, the up/down and down/up DW with the same $\varphi^0$ move in opposite directions (columns of the table Fig.~\ref{fig:Fig1}(b)). This leads to the possible expansion or collapse of magnetic domains under SOT. Usually, that is not possible with current drive in the usual cases: under SOT with DMI, the DW motion only depends on the chirality; whilst for the STT, its direction of motion only depends on $Q$~\cite{Shibata2011}. The DW motion also induces a tilt of its internal magnetization $\varphi$ which deviates from a pure N\'{e}el state (as shown Fig.~\ref{fig:Fig1}(b)). That tilting increases with the SOT magnitude, leading to larger effects at higher current densities.

Fig.~\ref{fig:Fig1}(c) shows the simulated DW dynamics at larger current density $J$ for a configuration $Q=1$ and $\varphi^0=0$. For each value of $J$, the DW velocity $v$ and angle $\varphi$ (black dots of the top and bottom panels) are extracted from the micromagnetic simulations after 20~ns of current injection. For low current, $v$ increases linearly with $J$ and $\varphi$ slightly deviates from its pure N\'{e}el state. When $\varphi$ reach the value of $\frac{\pi}{4}$ (blue dashed line), the DW velocity exhibits a maximum. For larger currents, the DW is pushed into its Bloch configuration. Since the SOT just acts on N\'{e}el DWs, the DW slows down. Above a critical current $J_\textrm{c}$ (red dashed line Fig.~\ref{fig:Fig1}(c)) the DW is locked into its pure Bloch configuration with $\varphi=\pm\frac{\pi}{2}$ and then $v=0$.

\subsection{Collective coordinate model} 

To understand the complex dynamics of such DW, it is convenient to use an analytical collective coordinate model that describes the DW dynamics with only a few parameters : $q$ and $\varphi$, as defined beforehand and shown Fig.~\ref{fig:Fig1}(a). The analytical integration of the LLG equation considering the rigid Bloch DW ansatz~\cite{Schryer1974} leads to coupled equations of motion~\cite{Mougin2007,Shibata2011,Thiaville2012}:
\begin{eqnarray}%
            \dot{q} &=& \frac{\Delta}{\alpha \tau_0}\left( \alpha^2 Q   j \cos \varphi + \frac{\sin 2\varphi}{2} \right), \label{eq:v}\\
        \dot{\varphi}  &=& \frac{1}{\tau_0}\left( Q  j \cos \varphi - \frac{\sin 2\varphi}{2}  \right), \label{eq:phip}
\end{eqnarray}
where $\Delta$ is the DW width (defined as positive). $\tau_0= \frac{1+\alpha^2}{\alpha \gamma_0 H_\textrm{KIP}}$ is a timescale associated with the DW dynamics, mainly depending on $H_\textrm{KIP}=\frac{2 K_\textrm{IP}}{\mu_0 M_\textrm{s}}$, the field associated to the shape anisotropy strength $K_\textrm{IP}$. The  dimensionless current ratio $j=\frac{J}{J_\textrm{c}}=\frac{\frac{\pi}{2}H_\textrm{SOT}}{\alpha H_\textrm{KIP}}$ results from the competition between the shape anisotropy, which tends to stabilize the N\'{e}el state of the DW and the SOT which tends to tilt the DW internal magnetization, accounted as an effective field $H_\textrm{SOT}$. 
Eq.~(\ref{eq:v}) and (\ref{eq:phip}) clearly show that $\varphi$ governs the DW dynamics and that $v$ depends on $\varphi$ roughly as $v\sim \sin 2\varphi$. Eq.~(\ref{eq:phip}) gives an implicit analytical expression of the time evolution of $\varphi$ such as :
\begin{equation}
    e^{-t/\tau_0} = \frac{\sqrt[\leftroot{0} \uproot{5} 2\left(Q j-1\right)]{\frac{1-\sin \phi}{1-\sin \phi^0}}}{\sqrt[\leftroot{0} \uproot{5} 2\left(Q j+1\right)]{\frac{1+\sin \phi}{1+\sin \phi^0}}\,\sqrt[\leftroot{0} \uproot{5} \left(j^2-1\right)]{\frac{Q j -\sin \phi}{Q j -\sin \phi^0}}},
   \label{eq:phiimpl}
\end{equation}
depending on $\varphi^0$, the internal angle at the considered initial time $t=0$.

For asymptotic steady state motion, the DW velocity and the internal angles are constant (\textit{i.e.} $\dot{q}$ and $\varphi$ take constant values that we denote $v^\infty$ and $\varphi^\infty$). Two steady state dynamical regimes are obtained in this case, separated by the critical current density $J_\textrm{c}$:
\begin{align}
v^\infty &=\begin{cases} \left(S_1\right)\;
\frac{\Delta}{\tau_0}   \frac{1+\alpha^2}{\alpha} j\sqrt{1-j^2}\; &\textrm{if} \; J	\leqslant J_\textrm{c}\;( j\leqslant1),\\
0\; &\textrm{if} \; J \geqslant J_\textrm{c}\;( j\geqslant 1),\end{cases} \label{eq:v_de_j_v} \\
                \varphi^\infty &=\begin{cases} (\widetilde{\varphi^0})+\left(S_1\right) \arcsin j \; &\textrm{if} \; J	\leqslant J_\textrm{c},\\
\left(S_2\right)\frac{\pi}{2}\; &\textrm{if} \; J	\geqslant J_\textrm{c}.\end{cases}
    \label{eq:v_de_j_phi}
\end{align}
$\left(S_1\right)$ and $\left(S_2\right)$ are equal to $\pm1$ and give the DW propagation and tilt angle directions depending on the initial state of the DW and on the current direction: $\left(S_1\right)=Q \sgn \left( \cos \varphi^0 \right)$ and $\left(S_2\right) = Q \sgn \left(J\right)$. $(\widetilde{\varphi^0})$ determines whether the DW is mainly left- or right-handed: $(\widetilde{\varphi^0})=0$ if $\varphi^0 \in ]-\frac{\pi}{2},\frac{\pi}{2}[$ and $=\pi$ if $\varphi^0 \in ]\frac{\pi}{2},\frac{3\pi}{2}[$. These expressions of $\left(S_1\right)$, $\left(S_2\right)$ and $(\widetilde{\varphi^0})$ reproduce well the results sketched in the table shown in Fig.~\ref{fig:Fig1}(b). These considerations also show that for $J< J_\textrm{c}$, $v^\infty$ and $\varphi^\infty$ depend of the initial condition $\varphi_0$ but not for $J \geqslant J_c$.

The variation of $v^\infty$ and $\varphi^\infty$ with $J$, plotted as yellow lines in Fig.~\ref{fig:Fig1}(c), also reproduce the simulated results with a very good quantitative agreement in the main part of the current range, far from $J_\textrm{c}$. Eq.~(\ref{eq:v_de_j_v}) and~(\ref{eq:v_de_j_phi}) also shows that a maximum velocity $\Delta\gamma_0 H_{\textrm{KIP}}$ is reached for $J=\frac{J_\textrm{c}}{\sqrt{2}}$ for which $\varphi^\infty=\pm\frac{\pi}{4}[\pi]$. Interestingly, that maximum only depends on the in-plane shape anisotropy but not directly on the SOT.

\subsection{Transient regime} 

\begin{figure}
    \includegraphics[width=8.5cm]{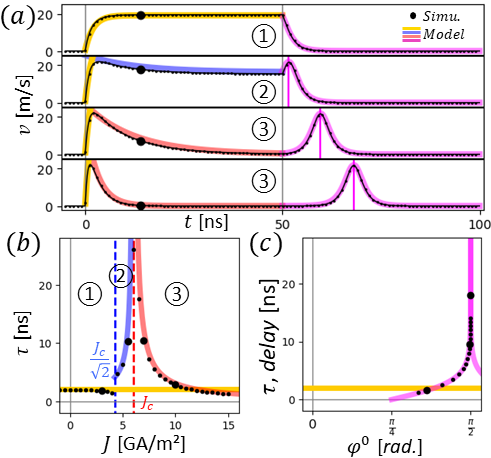}
    \caption{Transient regimes of the DW dynamics. (a) The different plots show the time evolution of the DW velocity during 50~ns current pulses of different magnitudes: case 1, $J=3$~GA/m$^2~< \frac{J_\textrm{c}}{\sqrt{2}}$; case 2, $J=5.5$~GA/m$^2~\in [\frac{J_\textrm{c}}{\sqrt{2}},J_\textrm{c}]$; case 3, $J=7$~GA/m$^2~> J_\textrm{c}$; and case 4, $J=10$~GA/m$^2~\gg J_\textrm{c}$. For each plot, black dots show the values from the numerical simulations. During the current injection (for $t<50$~ns), the colored lines are the exponential evolution with characteristic times $\tau$. After the current injection (for $t>50$~ns), the purple lines show the analytical model considering Eq.~(\ref{eq:v}) and (\ref{eq:phi_t_relax}). (b) Evolution of the characteristic time $\tau$ of the transient regime with $J$. Black dots correspond to the values of $\tau$ obtained with the exponential fits of the data in (a) and the colored lines show the analytical expression of $\tau$ for $t\approx0$ in yellow, and $t\rightarrow\infty$ in blue for $J \in [\frac{J_\textrm{c}}{\sqrt{2}},J_\textrm{c}]$ and in red for $J> J_\textrm{c}$. Blue and red dashed vertical lines correspond to $\frac{J_\textrm{c}}{\sqrt{2}}$ and $J_\textrm{c}$. (c) shows the DW motion delay depending on the DW angle when the current is turned off $\varphi^0$. The yellow horizontal line shows the characteristic time $\tau_0$ and the purple curve is the expected variation of the motion delay with $\varphi^0$. 
    }
    \label{fig:Fig2}
\end{figure}

The non-uniformity of the DW velocity with $\varphi$ has a strong effects in the transient regime. These are most easily visible around the critical current, where the simulated DW velocity does not seem to vanish immediately at $J\geq J_\textrm{c}$ as it is expected in the steady state motion. Fig.~\ref{fig:Fig2}(a) shows the time evolution of the DW velocity during a 50~ns current pulse of different magnitudes for an initial DW configuration $Q=1$ and $\varphi^0=0$. First, the loading time, just after the current onset is discussed. The relaxation dynamics is described in the following.

For $J<\frac{J_\textrm{c}}{\sqrt{2}}$ (case 1), the velocity, as well the internal angle (not shown here) tend continuously to the steady state values, as expected for classical DW motion. For $J>\frac{J_\textrm{c}}{\sqrt{2}}$ (case 2), the transient regime is divided into two parts: first the DW accelerates up to a maximum of velocity with a characteristic time of a few ns and then slows down, tending towards the steady state values with a much longer characteristic time. These two parts are due to the continuous evolution of $\varphi$ from $\varphi^0=0$ to the steady state value $\varphi^\infty$ which implies to cross $\pm\frac{\pi}{4}[\pi]$ where the velocity is at a maximum. The characteristic time of that velocity decay is much larger, up to tens of ns. For $J>J_\textrm{c}$, these two parts of the transient regime still exist. Even if the asymptotic velocity is zero, the DW moves for a significant amount of time before being locked into its Bloch configuration. That explains why the velocities extracted from the simulations (after 20~ns of current injection; in bold Fig.~\ref{fig:Fig2}(a)) shown Fig.~\ref{fig:Fig1}(c) are slightly above zero at and above $J_\textrm{c}$.

The variation of this characteristic time can be obtained by linearising Eq.~(\ref{eq:phip}) for $t\approx0$ around $\varphi^{0}$ or $t\rightarrow\infty$ around  $\varphi^{\infty}$. We can obtain that $\dot{\varphi}\approx\dot{\varphi}\left(\varphi^{0(\infty)}\right)-\frac{\varphi-\varphi^{0(\infty)} }{\tau}$ with $\dot{\varphi}\left(\varphi^{\infty}\right)\rightarrow0$ and so
\begin{equation}
    \tau = \frac{\tau_0}{  Q j\sin \varphi^{0(\infty)}  + \cos 2 \varphi^{0(\infty)}},
   \label{eq:tau}
\end{equation}
the characteristic time of the $\varphi$ variations at the beginning or at the end of the transient regime (i.e. for $\varphi\approx\varphi^{0}$ or $\varphi^{\infty}$). Thus, Eq.~(\ref{eq:tau}) captures the two different parts of the transient regime. For the first one, $\varphi^0=0$ or $\pi$ gives $\tau=\tau_0$ for any currents 
and $v_{(t\approx0)} \propto \left(\varphi_{(t\approx0)}-\varphi^0\right) \propto \left(1-e^{-t/\tau_0}\right)$. For the second part, when it exists (i.e. for $J> \frac{J_\textrm{c}}{\sqrt{2}}$), the characteristic time depends on $J$ explicitly and implicitly through $\varphi^\infty_{(J)}$ in Eq.~(\ref{eq:tau}). For $J\in ]\frac{J_\textrm{c}}{\sqrt{2}},J_\textrm{c}[$, $\tau_{(J)}= \frac{\tau_0}{1-j^2}$ and for $J>Jc$, $\tau_{(J)}=\frac{\tau_0}{j-1}$. For each case, 
$ \left(\varphi_{(t\approx\infty)}-\varphi^\infty\right) \propto \left(1-e^{-t/\tau}\right)$ and $\left(v_{(t\approx\infty)}-v^\infty\right) \propto e^{-t/\tau}$.
The colored lines in the Fig.~\ref{fig:Fig2}(a) show the exponential decays of $v$, with the extracted characteristic times shown as black dots in Fig.~\ref{fig:Fig2}(b). There are reproduced well by the analytical $\tau$ \textit{versus} $J$ curves (yellow : $J<\frac{J_\textrm{c}}{\sqrt{2}}$, blue : $J\in ]\frac{J_\textrm{c}}{\sqrt{2}},J_\textrm{c}[$ and red : $J>J_\textrm{c}$). This dependence of $\tau$ on the strength of the drive $J$ leads to non-trivial non-Newtonian dynamics. $\tau$ even diverges at $J_\textrm{c}$, which means that the internal angle tends to $\pm\frac{\pi}{2}$ but never reaches this value at this critical current. Conversely, for $J>J_\textrm{c}$, $\tau$ is finite, meaning that the DW reaches its pure Bloch configuration ($\varphi =\pm\frac{\pi}{2}$).

When the current is turned off, the DW relaxes in its N\'{e}el configuration (as shown Fig.~\ref{fig:Fig2}(a)) with $\varphi^\infty_{(J=0)}=0$ or $\pi$ depending on the internal DW angle at this time $\varphi^0$. For $J=0$, Eq.~(\ref{eq:phip}) gives:
\begin{equation}
    \varphi(t) = (\widetilde{\varphi^0})+\textrm{arccot}\left( \textrm{cot}\left(\varphi^0\right) e^{\sfrac{t}{\tau_0}}\right),
   \label{eq:phi_t_relax}
\end{equation}
with $(\widetilde{\varphi^0})=0$ or $\pi$ defined as before. The relaxation occurs with a characteristic time $\tau_0$, independent of $\varphi^0$. If the DW crosses $\varphi=\pm\frac{\pi}{4}[\pi]$ during its relaxation, it exhibits a velocity peak after a delay of $-\tau_0\,\textrm{log}\left| \textrm{cot}\left( \varphi^0\right)  \right|$ as visible Fig.~\ref{fig:Fig2}(a). Fig.~\ref{fig:Fig2}(c) shows the evolution of the motion delay whenever it exists \textit{versus} the DW angle at the end of the pulse ($\varphi^0$ in Eq.~(\ref{eq:phi_t_relax})).

The velocity overshoot of the DW for $J>\frac{J_\textrm{c}}{\sqrt{2}}$ might be very convenient for getting out eventual pinning centers as it is the case for other driving forces~\cite{Fukami2013,Hayashi2008} however, here, this property is intrinsic to the DW dynamics. Such spiky and delayed DW dynamics also makes echoes properties explored for brain-inspired computing~\cite{Yue2019,Blachowicz2020,Markovic2020}.




\subsection{Brownian motion under large current pulses} 

\begin{figure}
    \includegraphics[width=8cm]{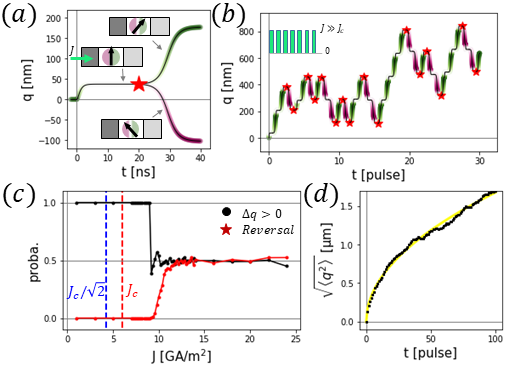}
    \caption{ Brownian motion of the DW under large current pulses: (a) shows the reversal process of the DW at the end of a current pulse with a magnitude larger than $J_\textrm{c}$. The evolution of the DW position is calculated by numerical integration of the Eq.~(\ref{eq:v}) and(\ref{eq:phip}) (black line) for a pulse of 20~ns width and 20 GA/m$^2$ magnitude. The colored highlight indicate if $\varphi\in]-\frac{\pi}{2},\frac{\pi}{2}[$ in green or  $\varphi\in]\frac{\pi}{2},\frac{3\pi}{2}[$ in purple. The insets sketch the DW state before and after the end of the pulse corresponding to a bifurcation point (indicated as a red star). (b) shows the DW under a series of 20 GA/m$^2$ 20~ns pulses with 20~ns delays (as indicated in the inset). To capture the infinitesimal changes of $\varphi$, at the end of each pulse $\varphi$ is deviated by an infinitesimal quantity obeying a normal distribution with a standard deviation of 0.1$^\circ$. The red stars indicate when the DW reversal occurs. (c) shows the probability of the DW motion in the positive direction (black curve) and of the DW reversal (red curve) calculated during 1,000 pulses for different current densities. (d) shows the square root evolution of the mean squared displacement $\left<q^2\right>$ with the number of pulses averaged over 200 series of 100 pulses. The yellow line is the linear fit $\left<q^2\right>\propto t$.}
    \label{fig:Fig6}
\end{figure}

It is interesting to note that $\varphi^\infty$ implicitly depends on $\varphi^0$ according to Eq.~(\ref{eq:v_de_j_phi}). Thus, the angles $\varphi^0=\pm\frac{\pi}{2}$ are bifurcation points where the asymptotic angles $\varphi^\infty$ towards which $\varphi$ will tend are completely different for $\varphi^0 = \pm \frac{\pi}{2} \pm 0^+$. In particular, infinitesimal changes lead to opposite propagation directions of the DW. However, for large current ($J\gg J_\textrm{c}$), required to reach the reversal points $\varphi=\pm\frac{\pi}{2}$, the SOT tends to keep the DW in that Bloch configuration (according to Eq.~(\ref{eq:v_de_j_phi}) and shown Fig.~\ref{fig:Fig1}). To achieve the reversal of the DW propagation taking advantage of these bifurcation points, we propose to use large current pulses, with magnitudes larger than $J_c$. The SOT pulse sets the DW in its Bloch state ($\varphi=\pm\frac{\pi}{2}$) and after the pulse, the DW relaxes into one of its N\'{e}el state $\varphi^\infty_{(J=0)}=0$ or $\pi$ randomly, inducing DW motions. This internal dynamics is shown Fig.~\ref{fig:Fig6}(a) for a 20~ns pulse of 15~GA/m$^2$ on a DW with an initial state $Q=1$ and $\varphi$=0. The insets and the colored highlight indicate if $\varphi\in]-\frac{\pi}{2},\frac{\pi}{2}[$ in green or  $\varphi\in]\frac{\pi}{2},\frac{3\pi}{2}[$ in purple. After the bifurcation point (shown as a red star) the DW relaxes in one of its N\'{e}el state driving the DW in opposite directions.

As the infinitesimal change of $\varphi$ just plays a role at the end of the pulse, it is possible to mimic the random reversals of the DW propagation during a large number of pulse by considering a small random deviation of the DW internal angle at the end of each pulse. In practice, this is naturally achieved by the thermal fluctuations or the inhomogeneities of the system. The Fig.~\ref{fig:Fig6}(b) shows the DW displacements and internal angle under a series of 20~GA/m$^2$ pulses with 20~ns widths and 20~ns delays. At the end of each pulses $\varphi$ is deviated by a quantity following a normal distribution with a standard deviation of $0.1^\textrm{o}$. The red stars indicate if the DW reversal occurs.

Fig.~\ref{fig:Fig6}(c) shows the probability of the DW displacement in the positive direction (black curve) and the probability of the propagation reversal (red curve) \textit{versus} the current density for 20~ns pulses. Above a critical current larger than $J_\textrm{c}$, the DW displacement in one or the other direction is completely equiprobable even if the reversal probability is gradually increasing. During this transition, the DW reversal is quite rare and the DW moves in the same direction for a large number of pulses. After this transition, the reversal is completely random and the DW motion obeys the Brownian law. Fig.~\ref{fig:Fig6}(d) shows the expected square root evolution of the mean squared DW displacement with the number of pulses~\cite{Toda1958} $\left<q^2\right>=2Dt$ with a diffusivity $D=1.4\times10^{-2}$ $\upmu$m$^2$/20~ns pulse. The current pulse magnitude above which the DW displacement is Brownian decreases with the pulse duration and with the standard deviation of the $\varphi$ deviation as these two quantities promote the DW reversal. This Brownian dynamics  does not depends on the current sequence and can be also obtained for a square- or a sine-wave current with large amplitude (not shown here). In the following, the reversal of a moving DW under continuous currents are considered.

\subsection{Propagation reversal under perpendicular field} 

As for a DW motion driven by continuous current $\dot{\varphi}_{(\varphi=\pm\frac{\pi}{2})}=0$ (according to Eq.~(\ref{eq:phip})), the controlled reversal of its propagation requires additional torques acting on $\varphi$. This could be achieved by using an in-plane field or field-like torques. However, this will not allow repeatably reversing the propagation without changing the polarity of these torques. This implies knowing the DW state at any moment to reverse its propagation. On the other hand, the use of an out-of-plane field $H_z$ induces a torque on the magnetization independent of $\varphi$ (only depending on $Q$). In a presence of $H_z$ Eq.~(\ref{eq:v}) and~(\ref{eq:phip}) become:
\begin{eqnarray}%
            \dot{q} &=& \frac{\Delta}{\alpha \tau_0 }\left( Q \alpha^2  \left[ j \cos \varphi +  h_z \right] + \frac{\sin 2\varphi}{2} \right), \label{eq:qpwithHz}\\
        \dot{\varphi}  &=& \frac{1}{\tau_0}\left( Q  \left[ j \cos \varphi +  h_z \right] - \frac{\sin 2\varphi}{2}  \right), \label{eq:phipwithHz}
\end{eqnarray}
with the dimensionless out-of-plane field $ h_z = \frac{H_z}{\alpha H_\textrm{KIP}}$ corresponding to the ratio between $H_z$ which tends to rotate $\varphi$ in the $Q H_z$ direction (whatever the value of $\varphi$) and $H_\textrm{KIP}$ which tends to keep the DW in its N\'{e}el state. In particular, when the field is applied, $\dot{\varphi}_{(\varphi=\pm\frac{\pi}{2},H_z\neq0)}\neq0$ which allows for overcoming the reversal limits $\varphi=\pm\frac{\pi}{2}$. Note that even if $v$ changes while $H_z$ is applied, the proper reversal is governed by the DW state when the field is turned-off. 

\begin{figure}
    \includegraphics[width=8cm]{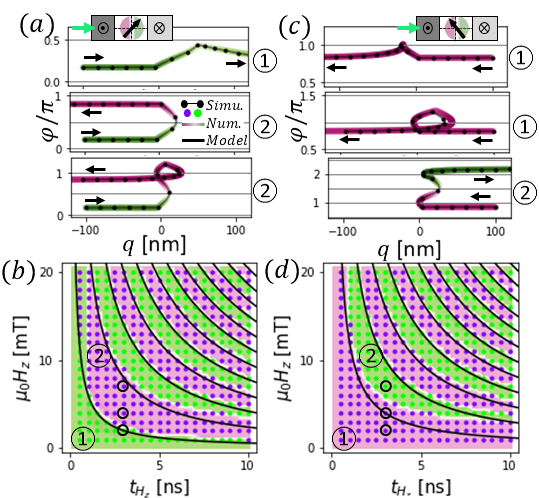}
    \caption{Propagation reversal under $H_z$: (a) and (b) show the propagation reversal of a DW driven by continuous SOT (with $J=3$~GA/m$^2$) under out-of-plane field pulses in the ($q$,$\varphi$) space for $Q=1$ and respectively, $\varphi^0\in]-\frac{\pi}{2},\frac{\pi}{2}[$ (a) and $\varphi^0\in]\frac{\pi}{2},\frac{3\pi}{2}[$ (b). For both (a) and (b), pulse width of duration $t_{H_z}=3$~ns and magnitude $\mu_0 H_z=$2, 4 and 7~mT and are applied for, respectively, the cases 1, 2, and 3. Black dots and lines correspond to the numerical simulations. The spacing between dots corresponds to a 1~ns time interval. The colored lines correspond to the numerical integration of the $q$-$\varphi$ model given by Eq.~(\ref{eq:qpwithHz}) and (\ref{eq:phipwithHz}). The color shows if $\varphi\in]-\frac{\pi}{2},\frac{\pi}{2}[$ in green, or if $\varphi\in]\frac{\pi}{2},\frac{3\pi}{2}[$ in purple, as indicated in the insets. The black arrows indicate the DW propagation directions.
    (c) and (d) show the reversal phase diagram in the $(H_z,t_{H_z})$ space for a DW driven by $J=3$~GA/m$^2$. Colored dots indicate the final state of the DW after 10~ns relaxation following the field pulse obtained by the micromagnetic simulations: $\varphi\in]-\frac{\pi}{2},\frac{\pi}{2}[$ in green and $\varphi\in]\frac{\pi}{2},\frac{3\pi}{2}[$ in purple. The black circles correspond to the configurations shown in (a) and (c). The background color shows the same results but obtained by the numerical integration of the $q$-$\varphi$ model (Eq.~(\ref{eq:qpwithHz}) and (\ref{eq:phipwithHz})). Black lines show the analytical hyperbolas, which separate the reversal regions obtained with the linearized model.}
    \label{fig:Fig3}
\end{figure}

Fig.~\ref{fig:Fig3}(a) and (b) show some examples of propagation reversal driven by field pulses for pulse width of $t_{H_z}=3$~ns and different magnitudes ($H_z>0$) plotted in the $(q,\varphi)$ space for $Q=1$. For every $(H_z,t_{H_z})$ pulse, the DW stabilizes itself in its steady state under a continuous current of $3$~GA/m$^2$ (during 5~ns), a pulse of field is applied during $t_{H_z}$, and the DW relaxes in its new state during 5~ns. To avoid the initial transient regimes, we consider $\varphi^0$ as $\varphi^0=\varphi^\infty\approx0.2\pi\in]-\frac{\pi}{2},\frac{\pi}{2}[$ or $\varphi^0=\varphi^\infty\approx0.8\pi\in]\frac{\pi}{2},\frac{3\pi}{2}[$ for (a) or (b), respectively. The black arrows indicate the propagation direction of the DW. In both (a) and (b), the black dots correspond to the evolution of the DW state obtained from the micromagnetic simulations with the space between them corresponding to 1~ns time intervals. The colored lines correspond to the numerical integration of the $q$-$\varphi$ model accounting a field pulse (given by Eq.~(\ref{eq:v}) and (\ref{eq:phip}) in absence of the field and Eq.~(\ref{eq:qpwithHz}) and (\ref{eq:phipwithHz}) during the field pulse). The color of these lines depends on $\varphi$: green for $\varphi\in]-\frac{\pi}{2},\frac{\pi}{2}[$ and purple for $\varphi\in]\frac{\pi}{2},\frac{3\pi}{2}[$. The plots shown in  Fig.~\ref{fig:Fig3}(a) and (b) clearly demonstrate that the applied field can be used to overcome the reversal limit which flips the propagation direction of the DW after the field for, possibly, the two states of the DW.

This field-induced reversal strongly depends on the magnitude and width of the field pulse. To handle that dependence, the above simulation has been repeated for different $(H_z,t_{H_z})$. Fig.~\ref{fig:Fig3}(c) and~(d) show the final state of the DW after the field pulse for each $(H_z,t_{H_z})$ (obtained with the micromagnetic simulations - dots - and  by numerically integrating the $q$-$\varphi$ model - background color). The configurations detailed in (a) and (b) are circled in black in (c) and (d). To obtain the reversal conditions, Eq.~(\ref{eq:phipwithHz}) can be linearized by considering the effect of the out-of-plane field on $\varphi$ is much larger than the combined effects of the SOT and the in-plane anisotropy giving 
$\varphi\approx\varphi^0 +\frac{Q h_z}{\tau_0}t$. The conditions for a DW reversal can thus be defined as  $\left(\frac{4n+1}{2}\pi-\varphi^\infty\right)< Q h_z \frac{ t_{H_z}}{\tau_0}  < \left(\frac{4n+3}{2}\pi-\varphi^\infty\right)$ for $\varphi^0\in]-\frac{\pi}{2},\frac{\pi}{2}[$ (the cases shown in Fig.~\ref{fig:Fig3} (a) and (c)), and $\left(\frac{4n-1}{2}\pi+\varphi^\infty\right)< Q h_z \frac{ t_{H_z}}{\tau_0} < \left(\frac{4n-3}{2}\pi+\varphi^\infty\right)$ for $\varphi^0\in]\frac{\pi}{2},\frac{3\pi}{2}[$ (the cases shown in Fig.~\ref{fig:Fig3} (b) and (d)). $n$ corresponds to the number of flips in multiple flipping processes.

The analytic hyperbolic lines delimiting these regions in the $(H_z,t_{H_z})$ space are plotted in black in Fig.~\ref{fig:Fig3}(c) and (d). They match well to the simulated cases except for low fields and low pulse widths where the $\dot{\varphi}\sim H_z$ assumption is not well justified. For larger $H_z$ and $t_{H_z}$, the regions where the DW propagation is reversed are more and more condensed, which is equivalent to a quasirandom flipping process (similar to the coin reversal in a heads-or-tails process~\cite{Diaconis2007}).

Additionally, these reversal regions are not exactly the same for $\varphi^0\in]-\frac{\pi}{2},\frac{\pi}{2}[$  and  $\varphi^0\in]\frac{\pi}{2},\frac{3\pi}{2}[$. Thus, it is possible to reverse only the DW appertaining to one of these cases, setting all the DW with the same $\varphi$, as shown in the second panel in Fig.~\ref{fig:Fig3}(a) and (b). At the opposite, a pulse with conditions that correspond to a reversal for the two cases flips all the DWs, independently of their initial angle, as shown third panel in Fig.~\ref{fig:Fig3}(a) and (b). The shift between the reversal regions for $\varphi^0\in]-\frac{\pi}{2},\frac{\pi}{2}[$ and $\varphi^0\in]\frac{\pi}{2},\frac{3\pi}{2}[$ increases with the current density as $\varphi^\infty$ is much closer to the reversal point for one type of DW than for the other (as visible in the table Fig.~\ref{fig:Fig1}(b)).

\subsection{Step in anisotropy} 

Even if an out-of-plane field allows the propagation reversal, its effects are strongly dependent on the pulse width and magnitude, and significantly disturb the DW propagation with multiple reversals. To achieve similar effects in a much more controlled way, it is possible to consider a spatial variation of the out-of-plane anisotropy $K_{(x)}$ (with $x$ the position along the wire). That can be directly associated with an effective local out-of-plane field $H^K_{z(q)} = -\frac{\Delta}{\mu_0 M_s}\partial_q \tilde{K}_{(q)}$ where $\tilde{K}_{(q)}=\frac{\int K_{(x)} (1-m_{z(x,q)}^2) dx}{\int (1-m_{z(x,q)}^2) dx}$ is the effective variation of the out-of-plane anisotropy with $q$, the position of the DW along the magnetic wire. Here, $m_z$ is the $z$-component of the local unit magnetization vector $\hat{\bm{m}}$. This field vanishes where the anisotropy is uniform and it has a maximum where $K$ changes most rapidly with position $x$. 

\begin{figure}
    \includegraphics[width=8cm]{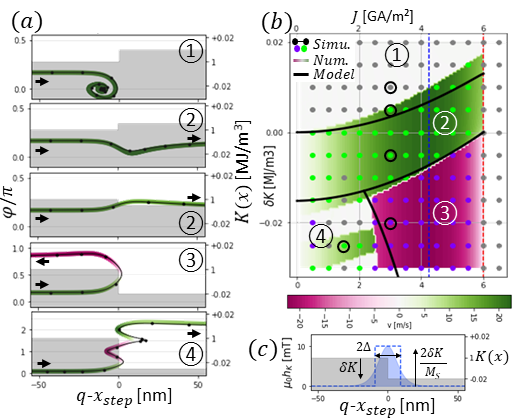}
    \caption{Propagation reversal at an anisotropy step: (a) shows the propagation reversal of a DW at the anisotropy step at the position $x_\textrm{step}$ in the $(q,\varphi)$ space for $\varphi^0=\varphi^\infty\approx0.2\pi$ and $Q=1$ and for $(J,\delta K) =(3,0.01)$, $(3,0.005)$, $(3,-0.005)$, $(3,-0.02)$, and $(1.5,-0.025)$ in units of (GA/m$^2$, MJ/m$^3$). The gray shape in the background (relative to the right axis) shows the x-variation of the anisotropy. Black dots and lines correspond to the numerical simulations with a spacing between the dots corresponding to a 1~ns time interval. The colored lines correspond to the numerical integration of the $q$-$\varphi$ model considering the anisotropy step as a local field $H_z^K(q)$. Its color shows whether $\varphi\in]-\frac{\pi}{2},\frac{\pi}{2}[$ in green and $\varphi\in]\frac{\pi}{2},\frac{3\pi}{2}[$ in purple. The black arrows indicate the DW propagation direction. (b) shows the reversal phase diagram in the $(J,\delta K)$ space. Colored dots indicate the final state of the DW after its interaction with the anisotropy step obtained by the numerical simulations: transmission in green, stoppage in gray, and reflected in purple. The black circles correspond to the configurations shown in (a). The background color show the same results but obtained by the numerical integration of the $q$-$\varphi$ model. Black lines show the frontiers between the different behaviors considering an approximated square field as discussed in the main text. (c) shows a sketch of the different approximations of the effective field $H_z^K(q)$: the gray shape in the background (relative to the right axis) shows the variation of the anisotropy, the blue curve (relative to the left axis) shows the variation of $H_z^K(q)$ (by assuming a rigid DW) and the dashed blue curve shows the square approximation of $H_z^K(q)$ used for the simplified $q$-$\varphi$ model.}
    \label{fig:Fig4}
\end{figure}

Fig.~\ref{fig:Fig4}(a) shows the effect of a step of anisotropy of magnitude $\delta K$ at the position $x_\textrm{step}$ on a moving DW with $Q=1$ and $\varphi^0 = \varphi^\infty$ for different $J$ and different step magnitudes. The evolution of the DW dynamics is represented in the $(q,\varphi)$ space and the arrows indicate the DW propagation directions. The black dots correspond to the simulated evolution of the DW dynamics with a time interval between them of 1~ns. The gray curve in the background shows the anisotropy variation along the wire. As visible in the different cases, the DW can be either stopped (case 1), transmitted (case 2 or 4), or reflected (case 3), depending on $J$ and $\delta K$.

As for a regular out-of-plane field, $H_{\delta K}$ acts on $\varphi$ as described by the Eq.~(\ref{eq:phipwithHz}), but here the strength of the field experienced by the DW depends on its position along the wire $q$. For high positive $\delta K$, if $H_{z(q)}^K\propto-\delta K$ pushes $\varphi$ down to zero, the DW propagation is reversed (since $v \sim \sin 2\varphi$) and the field experienced by the DW decreases. That pushes the DW back to its steady state with $\varphi^\infty >0$, reversing again its propagation direction. The converging repetition of that process leads to the pinning of the DW before the anisotropy step as shown in the first panel of Fig.~\ref{fig:Fig4}(a) (case 1). 

If $\varphi$ does not cross zero while the DW cross the region where the anisotropy changes, the DW propagation is not reversed and the DW is transmitted as shown in the second panel of Fig.~\ref{fig:Fig4}(a) (case 2). That occurs for lower $\delta K$, where $\varphi$ is less affected by the anisotropy step, and for larger $J$ where $\varphi^\infty$ is farther from $0$ and the DW is faster, experiencing the effective field for a shorter period of time. For negative $\delta K$, $H_z^K$ pushes $\varphi$ farther from zero. For low negative $\delta K$ and low $J$, the DW crosses the anisotropy step without crossing the reversal limit $\varphi=\frac{\pi}{2}$, and so it is transmitted, as shown in the third panel (case 2) in Fig.~\ref{fig:Fig4}(a). 

For larger $J$, $\varphi^\infty$ is larger and even a small $\delta K$ allows the DW to cross this reversal limit. The DW moves then farther from the anisotropy step and experiences $H_z^K$ less and less. The DW is therefore reflected by the anisotropy step as shown in the fourth panel of Fig.~\ref{fig:Fig4}(a), case 3. 

For large and negative $\delta K$, the proximity of the anisotropy step can lead to multiple reversals of $\varphi$. If the current is too small, the DW is either reflected (for an odd number of reversals) or transmitted (for an even number of reversals; as shown in the fifth panel of Fig.~\ref{fig:Fig4}(a), case 4). For larger current, the DW reverses just one time and the DW is reflected similarly to case 3. These different effects of the anisotropy step are summarized in Fig.~\ref{fig:Fig3}(b). The colored points corresponds to the $(J,\delta K)$ configurations where the DW is either stopped (gray points), transmitted (green points), or reflected (purple points).

By assuming $\delta K$ is quite small compared to $K$, the DW width $\Delta$ is considered as constant, independent of the anisotropy step \cite{Schryer1974}. A sharp step of anisotropy at a position $x_\textrm{step}$ with a magnitude of $\delta K$, is thus directly associated to an out-of-plane field $H_{z\left(q-x_\textrm{step}\right)}^K = -\frac{\delta K}{2\mu_0 M_\textrm{s}}  \cosh^{-2} \left(Q \frac{q-x_\textrm{step}}{\Delta}\right)$, as plotted in Fig.~\ref{fig:Fig4}(c). By numerically integrating the coupled space and time evolution of $q$ and $\varphi$ in Eq.~(\ref{eq:qpwithHz}) and (\ref{eq:phipwithHz}), considering the above $H_{z\left(q-x_\textrm{step}\right)}^K$, it is possible to reproduce the simulated results with a very good agreement. The lines in Fig.~\ref{fig:Fig4}(a) show the time evolution of $q$ and $\varphi$ by considering this effective local field with the colors indicating the DW state: $\varphi\in]-\frac{\pi}{2},\frac{\pi}{2}[$ (in green) and  $\varphi\in]\frac{\pi}{2},\frac{3\pi}{2}[$ (in purple). The background colors in Fig.~\ref{fig:Fig4}(b) show the final velocity of the DW obtained with this approach: stopped in gray, transmitted in green, and reflected in purple.

The DW reversal conditions can even be obtained by considering $H_{\delta K}$ as a uniform field of magnitude $H_z^K=-\frac{\delta K}{ \mu_0 M_\textrm{s}}$ for $q\in [ x_\textrm{step}-\Delta, x_{step}+\Delta]$ and 0 outside this region (as plotted Fig.~\ref{fig:Fig4}(c)). Therefore, the reflection, transmission, or pinning of the DW is given by the DW final state for $q= x_\textrm{step}\pm\Delta$ after the DW interaction with the anisotropy step. Since $H_\textrm{SOT} \ll H_z^K$ and $\alpha H_\textrm{KIP} \ll H_z^K$, Eq.~(\ref{eq:qpwithHz}) and (\ref{eq:phipwithHz}) can be linearized  such as : $\dot{q} \approx \frac{\Delta}{\alpha \tau_0 }\left( - Q \alpha \frac{\delta K}{2 K_{IP}}  + \frac{\sin 2\varphi}{2} \right) $ and $\dot{\varphi}  \approx - \frac{Q \delta K }{2\alpha K_{IP}\tau_0}$. The frontiers between the different regions of the diagram in Fig.~\ref{fig:Fig4}(b) can thus be extracted. For $\varphi_{\left(q = x_\textrm{step}-\Delta \right)} = \varphi^\infty\in]-\frac{\pi}{2},\frac{\pi}{2}[$, the frontier between the regions 1 and 2 in Fig.~\ref{fig:Fig4}(b) is given by $\varphi_{\left(q= x_{step}+\Delta \right)} \leq 0$. This gives $\delta K_{[1/2]} = Q K_\textrm{IP} \left(\frac{\sin^2 \varphi^\infty}{2+\alpha \varphi^\infty}\right)$. The frontiers between region 2 and regions 3 and 4 is given by $\varphi_{\left(q = x_\textrm{step}+\Delta \right)} \geq \frac{\pi}{2}$ which leads to $\delta K_{[2/3]} = \delta K_{[2/4]} = -Q K_\textrm{IP} \left(\frac{ \cos^2 \varphi^\infty}{2-\alpha\left( \frac{\pi}{2}- \varphi^\infty\right)} \right)$. The frontier between regions 3 and 4 is given by two conditions:  $\varphi_{\left(q = x_\textrm{step}+\Delta \right)} \geq  \frac{\pi}{2}$ as before, but also $\varphi_{\left(q = x_\textrm{step}-\Delta \right)} \leq \pi$ corresponding to the two reversal process case. This gives $\delta K_{[3/4]} = -Q K_\textrm{IP} \left(\frac{\sin^2 \varphi^\infty}{\alpha\left( \pi- \varphi^\infty\right)} \right)$. These frontiers are shown by black lines in Fig.~\ref{fig:Fig4}(b) and match both the simulated and the numerical results with a good agreement. 
Since these calculations are based on the linearization of Eq.~(\ref{eq:phipwithHz}), they are more accurate if $\varphi$ only deviates slightly from $\varphi_{\left(q = x_\textrm{step}-\Delta \right)}=\varphi^\infty$, such as the frontier between 1 and 2 for low current and the one between 2 and 3 for $J$ close to $J_\textrm{c}$. This results show that by adapting the anisotropy step $\delta K$ to the current driving the DW, it is possible to select one of the processes. Such locally variable anisotropy can be controlled by using voltages \cite{Li2015,Lin2016}, and can lead to a three-state DW transistor.

\begin{figure}
    \includegraphics[width=8.5cm]{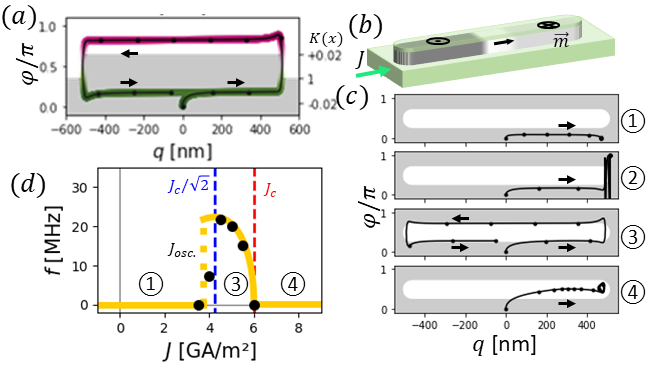}
    \caption{DW-based oscillators: (a) shows an example of the DW oscillations for a DW between two anisotropy steps (similar to the case 3 in Fig.~\ref{fig:Fig4}(a) with a continuous current of $J=3$~GA/m$^2$ and $\delta K = \pm$0.02~MJ/m$^3$ and for an initial state $Q=1$ and $\varphi^0=0$. Black dots, black lines, the gray profile, and colored lines have the same meanings as in Fig.~\ref{fig:Fig4}(a). (b) shows a sketch of a DW oscillator based on a finite magnetic wire where the continuous electrical current is injected in the bottom HM layer. (c) shows the different behaviors of the DW in such devices: DW stopped, destroyed or reflected by the track extremities, depending on the current density for respectively: $J= 1.5,\,3,\, 4.5,\, 7$ GA/m$^2$ in the ($q$,$\varphi$) space. The background indicates the shape of the finite magnetic track. (d) shows the variation of the DW oscillations frequency for different magnitude of the DC current extracted from (c). The yellow curve corresponds to the expected frequency of these oscillations $f=v/2l$ with $l$= 1$\mu$m above a certain current density $J_{osc.}$.}
    \label{fig:Fig5}
\end{figure}

In the case of a DW between two anisotropy steps corresponding to the region 3 of the diagram Fig.~\ref{fig:Fig4}(b), the DW will be reflected at each of the two steps leading to the oscillations of the DW propagation as shown Fig.~\ref{fig:Fig5}(a), analogous to confinement in a Fabry-Perot cavity.

\subsection{Finite magnetic track} 

A similar effect can be obtained at the extremities of a finite magnetic wire, as shown Fig.~\ref{fig:Fig5}(b) and (c). The black points in Fig.~\ref{fig:Fig5}(c) correspond to the simulated trajectory of the DW in $(q,\varphi)$ space and the background color corresponds to the shape of the magnetic element (visible Fig.~\ref{fig:Fig5}(b)). As previously, the extremities of the magnetic wire act on the DW internal angle $\varphi$. For low current, the DW stopped in its Bloch configuration (here $\varphi = 0$) just before the end of the wire, as shown in the first case of the  Fig.~\ref{fig:Fig5}(c). If the current increases, the DW goes off the end of the track before being in its Bloch state and the magnetic wire is then fully saturated as shown in the second case of the  Fig.~\ref{fig:Fig5}(c). For a sufficiently large current, the end-of-track effect pushes the DW across its reversal points before it leaves the wire and the DW is then reflected along the track. A similar process occurs at the other extremity and then the DW oscillates continuously from one to the other extremity, as in the third case of Fig.~\ref{fig:Fig5}(c). If $J>J_\textrm{c}$ and the DW is in the middle of the wire, far from the track extremities, the DW is stopped as expected. However, if the DW is close to the track end, it will be attracted and then stopped in its Bloch state after small oscillations of $\varphi$ around $\frac{\pi}{2}$ as in the fourth case of the  Fig.~\ref{fig:Fig5}(c).

The possibility to make DW oscillate in such small elements driven by a continues current is also a new way for small magnetic oscillators with a triangular wave time evolution, and for which the frequency $f=v/2l$, typically in the MHz range, is just governed by the DW velocity $v$ and the track length $l$. Fig.~\ref{fig:Fig5}(d) shows the variation of this frequency versus the magnitude of the current for the magnetic track shows in Fig.~\ref{fig:Fig5}(b). Interestingly, the non-uniform variation of $v$ with $J$ leads to a non-lineal variation of this frequency with the magnitude of the continuous current above a critical current $J_{osc.}$ above which the DW oscillates. The shape of the track extremity has only small effects on these results and just modifies $J_{osc.}$.

\section{Conclusion}

The possibility to stabilize N\'{e}el DWs with an achiral mechanism allows for the bistability of these magnetic textures, where both chiralities are energetically degenerate. This gives rise to new forms of SOT-driven DW dynamics. Two propagation directions for such DWs are now possible, only depending on their initial state. By increasing the current density, the DW exhibits a non-monotonic velocity law with a maximum of velocity and a critical current above which the DW is locked in its pure Bloch state and does not move. This non-uniformity of the velocity law also induces important transient regimes and out-of-equilibrium dynamics. Finally, the possibility of controlled or random reversal of the DW chirality, and therefore, of its propagation direction, has been demonstrated under certain conditions (large current pulses, out-of-plane field, steps of magnetic anisotropy or the extremities of finite length magnetic tracks). Of particular note are the ability to reflect or transmit DWs at a step in the anisotropy $K$, with the potential for control through voltage-controlled magnetic anisotropy, and new forms of DW oscillator in which a DW is continuously reflected between anisotropy seps or the ends of a magnetic nanowire. All these completely new behaviors, simulated numerically, have been perfectly reproduced with a suitable analytical model. All these new dynamics offer very interesting possibilities for future DWs-based devices that can possibly mix storage and processing.

\begin{acknowledgements}
This project has received funding from the EPSRC  under grant number EP/T006803/1 and the European Union’s Horizon 2020 research and innovation program under the Marie Sklodowska-Curie grant agreement No 750147. K.J.A.F. acknowledges support from the Jane and Aatos Erkko Foundation. The main part of these numerical simulations were performed on MAGI, the computing platform of the University Sorbonne Paris Nord, managed by Nicolas Greneche. We thank Yves Roussign\'{e} (from the LSPM) for fruitful discussions.

Data and code associated with this work are available on the University of Leeds Data Repository at https://doi.org/TBC.
\end{acknowledgements}

\bibliography{biblio}

\begin{thebibliography}{47}%
\makeatletter
\providecommand \@ifxundefined [1]{%
 \@ifx{#1\undefined}
}%
\providecommand \@ifnum [1]{%
 \ifnum #1\expandafter \@firstoftwo
 \else \expandafter \@secondoftwo
 \fi
}%
\providecommand \@ifx [1]{%
 \ifx #1\expandafter \@firstoftwo
 \else \expandafter \@secondoftwo
 \fi
}%
\providecommand \natexlab [1]{#1}%
\providecommand \enquote  [1]{``#1''}%
\providecommand \bibnamefont  [1]{#1}%
\providecommand \bibfnamefont [1]{#1}%
\providecommand \citenamefont [1]{#1}%
\providecommand \href@noop [0]{\@secondoftwo}%
\providecommand \href [0]{\begingroup \@sanitize@url \@href}%
\providecommand \@href[1]{\@@startlink{#1}\@@href}%
\providecommand \@@href[1]{\endgroup#1\@@endlink}%
\providecommand \@sanitize@url [0]{\catcode `\\12\catcode `\$12\catcode
  `\&12\catcode `\#12\catcode `\^12\catcode `\_12\catcode `\%12\relax}%
\providecommand \@@startlink[1]{}%
\providecommand \@@endlink[0]{}%
\providecommand \url  [0]{\begingroup\@sanitize@url \@url }%
\providecommand \@url [1]{\endgroup\@href {#1}{\urlprefix }}%
\providecommand \urlprefix  [0]{URL }%
\providecommand \Eprint [0]{\href }%
\providecommand \doibase [0]{https://doi.org/}%
\providecommand \selectlanguage [0]{\@gobble}%
\providecommand \bibinfo  [0]{\@secondoftwo}%
\providecommand \bibfield  [0]{\@secondoftwo}%
\providecommand \translation [1]{[#1]}%
\providecommand \BibitemOpen [0]{}%
\providecommand \bibitemStop [0]{}%
\providecommand \bibitemNoStop [0]{.\EOS\space}%
\providecommand \EOS [0]{\spacefactor3000\relax}%
\providecommand \BibitemShut  [1]{\csname bibitem#1\endcsname}%
\let\auto@bib@innerbib\@empty
\bibitem [{\citenamefont {Kumar}\ \emph {et~al.}(2022)\citenamefont {Kumar},
  \citenamefont {Jin}, \citenamefont {Sbiaa}, \citenamefont {Kl{\"{a}}ui},
  \citenamefont {Bedanta}, \citenamefont {Fukami}, \citenamefont {Ravelosona},
  \citenamefont {Yang}, \citenamefont {Liu},\ and\ \citenamefont
  {Piramanayagam}}]{Kumar2022}%
  \BibitemOpen
  \bibfield  {author} {\bibinfo {author} {\bibfnamefont {D.}~\bibnamefont
  {Kumar}}, \bibinfo {author} {\bibfnamefont {T.}~\bibnamefont {Jin}}, \bibinfo
  {author} {\bibfnamefont {R.}~\bibnamefont {Sbiaa}}, \bibinfo {author}
  {\bibfnamefont {M.}~\bibnamefont {Kl{\"{a}}ui}}, \bibinfo {author}
  {\bibfnamefont {S.}~\bibnamefont {Bedanta}}, \bibinfo {author} {\bibfnamefont
  {S.}~\bibnamefont {Fukami}}, \bibinfo {author} {\bibfnamefont
  {D.}~\bibnamefont {Ravelosona}}, \bibinfo {author} {\bibfnamefont {S.-H.}\
  \bibnamefont {Yang}}, \bibinfo {author} {\bibfnamefont {X.}~\bibnamefont
  {Liu}},\ and\ \bibinfo {author} {\bibfnamefont {S.}~\bibnamefont
  {Piramanayagam}},\ }\bibfield  {title} {\bibinfo {title} {{Domain wall
  memory: Physics, materials, and devices}},\ }\href
  {https://doi.org/10.1016/j.physrep.2022.02.001} {\bibfield  {journal}
  {\bibinfo  {journal} {Phys. Rep.}\ }\textbf {\bibinfo {volume} {958}},\
  \bibinfo {pages} {1} (\bibinfo {year} {2022})}\BibitemShut {NoStop}%
\bibitem [{\citenamefont {Parkin}\ \emph {et~al.}(2008)\citenamefont {Parkin},
  \citenamefont {Hayashi},\ and\ \citenamefont {Thomas}}]{Parkin2008a}%
  \BibitemOpen
  \bibfield  {author} {\bibinfo {author} {\bibfnamefont {S.~S.~P.}\
  \bibnamefont {Parkin}}, \bibinfo {author} {\bibfnamefont {M.}~\bibnamefont
  {Hayashi}},\ and\ \bibinfo {author} {\bibfnamefont {L.}~\bibnamefont
  {Thomas}},\ }\bibfield  {title} {\bibinfo {title} {{Magnetic Domain-Wall
  Racetrack Memory}},\ }\href {https://doi.org/10.1126/science.1145799}
  {\bibfield  {journal} {\bibinfo  {journal} {Science}\ }\textbf {\bibinfo
  {volume} {320}},\ \bibinfo {pages} {190} (\bibinfo {year}
  {2008})}\BibitemShut {NoStop}%
\bibitem [{\citenamefont {Allwood}\ \emph {et~al.}(2005)\citenamefont
  {Allwood}, \citenamefont {Xiong}, \citenamefont {Faulkner}, \citenamefont
  {Atkinson}, \citenamefont {Petit},\ and\ \citenamefont
  {Cowburn}}]{Allwood2005a}%
  \BibitemOpen
  \bibfield  {author} {\bibinfo {author} {\bibfnamefont {D.~A.}\ \bibnamefont
  {Allwood}}, \bibinfo {author} {\bibfnamefont {G.}~\bibnamefont {Xiong}},
  \bibinfo {author} {\bibfnamefont {C.~C.}\ \bibnamefont {Faulkner}}, \bibinfo
  {author} {\bibfnamefont {D.}~\bibnamefont {Atkinson}}, \bibinfo {author}
  {\bibfnamefont {D.}~\bibnamefont {Petit}},\ and\ \bibinfo {author}
  {\bibfnamefont {R.~P.}\ \bibnamefont {Cowburn}},\ }\bibfield  {title}
  {\bibinfo {title} {{Magnetic Domain-Wall Logic}},\ }\href
  {https://doi.org/10.1126/science.1108813} {\bibfield  {journal} {\bibinfo
  {journal} {Science}\ }\textbf {\bibinfo {volume} {309}},\ \bibinfo {pages}
  {1688} (\bibinfo {year} {2005})}\BibitemShut {NoStop}%
\bibitem [{\citenamefont {Lin}\ \emph {et~al.}(2022)\citenamefont {Lin},
  \citenamefont {Xu}, \citenamefont {Wang}, \citenamefont {Liu}, \citenamefont
  {Zhao}, \citenamefont {Zhou}, \citenamefont {Luo}, \citenamefont {Song},
  \citenamefont {Yu},\ and\ \citenamefont {Xing}}]{Lin2022}%
  \BibitemOpen
  \bibfield  {author} {\bibinfo {author} {\bibfnamefont {H.}~\bibnamefont
  {Lin}}, \bibinfo {author} {\bibfnamefont {N.}~\bibnamefont {Xu}}, \bibinfo
  {author} {\bibfnamefont {D.}~\bibnamefont {Wang}}, \bibinfo {author}
  {\bibfnamefont {L.}~\bibnamefont {Liu}}, \bibinfo {author} {\bibfnamefont
  {X.}~\bibnamefont {Zhao}}, \bibinfo {author} {\bibfnamefont {Y.}~\bibnamefont
  {Zhou}}, \bibinfo {author} {\bibfnamefont {X.}~\bibnamefont {Luo}}, \bibinfo
  {author} {\bibfnamefont {C.}~\bibnamefont {Song}}, \bibinfo {author}
  {\bibfnamefont {G.}~\bibnamefont {Yu}},\ and\ \bibinfo {author}
  {\bibfnamefont {G.}~\bibnamefont {Xing}},\ }\bibfield  {title} {\bibinfo
  {title} {{Implementation of Highly Reliable and Energy‐Efficient
  Nonvolatile In‐Memory Computing using Multistate Domain Wall Spin–Orbit
  Torque Device}},\ }\href {https://doi.org/10.1002/aisy.202200028} {\bibfield
  {journal} {\bibinfo  {journal} {Advanced Intelligent Systems}\ }\textbf
  {\bibinfo {volume} {4}},\ \bibinfo {pages} {2200028} (\bibinfo {year}
  {2022})}\BibitemShut {NoStop}%
\bibitem [{\citenamefont {Luo}\ \emph {et~al.}(2019)\citenamefont {Luo},
  \citenamefont {Dao}, \citenamefont {Hrabec}, \citenamefont {Vijayakumar},
  \citenamefont {Kleibert}, \citenamefont {Baumgartner}, \citenamefont {Kirk},
  \citenamefont {Cui}, \citenamefont {Savchenko}, \citenamefont {Krishnaswamy},
  \citenamefont {Heyderman},\ and\ \citenamefont {Gambardella}}]{Luo2019}%
  \BibitemOpen
  \bibfield  {author} {\bibinfo {author} {\bibfnamefont {Z.}~\bibnamefont
  {Luo}}, \bibinfo {author} {\bibfnamefont {T.~P.}\ \bibnamefont {Dao}},
  \bibinfo {author} {\bibfnamefont {A.}~\bibnamefont {Hrabec}}, \bibinfo
  {author} {\bibfnamefont {J.}~\bibnamefont {Vijayakumar}}, \bibinfo {author}
  {\bibfnamefont {A.}~\bibnamefont {Kleibert}}, \bibinfo {author}
  {\bibfnamefont {M.}~\bibnamefont {Baumgartner}}, \bibinfo {author}
  {\bibfnamefont {E.}~\bibnamefont {Kirk}}, \bibinfo {author} {\bibfnamefont
  {J.}~\bibnamefont {Cui}}, \bibinfo {author} {\bibfnamefont {T.}~\bibnamefont
  {Savchenko}}, \bibinfo {author} {\bibfnamefont {G.}~\bibnamefont
  {Krishnaswamy}}, \bibinfo {author} {\bibfnamefont {L.~J.}\ \bibnamefont
  {Heyderman}},\ and\ \bibinfo {author} {\bibfnamefont {P.}~\bibnamefont
  {Gambardella}},\ }\bibfield  {title} {\bibinfo {title} {{Chirally coupled
  nanomagnets}},\ }\href {https://doi.org/10.1126/science.aau7913} {\bibfield
  {journal} {\bibinfo  {journal} {Science}\ }\textbf {\bibinfo {volume}
  {363}},\ \bibinfo {pages} {1435} (\bibinfo {year} {2019})}\BibitemShut
  {NoStop}%
\bibitem [{\citenamefont {Liu}\ \emph {et~al.}(2021)\citenamefont {Liu},
  \citenamefont {Xiao}, \citenamefont {Cui}, \citenamefont {Incorvia},
  \citenamefont {Bennett},\ and\ \citenamefont {Marinella}}]{Liu2021}%
  \BibitemOpen
  \bibfield  {author} {\bibinfo {author} {\bibfnamefont {S.}~\bibnamefont
  {Liu}}, \bibinfo {author} {\bibfnamefont {T.~P.}\ \bibnamefont {Xiao}},
  \bibinfo {author} {\bibfnamefont {C.}~\bibnamefont {Cui}}, \bibinfo {author}
  {\bibfnamefont {J.~A.~C.}\ \bibnamefont {Incorvia}}, \bibinfo {author}
  {\bibfnamefont {C.~H.}\ \bibnamefont {Bennett}},\ and\ \bibinfo {author}
  {\bibfnamefont {M.~J.}\ \bibnamefont {Marinella}},\ }\bibfield  {title}
  {\bibinfo {title} {{A domain wall-magnetic tunnel junction artificial synapse
  with notched geometry for accurate and efficient training of deep neural
  networks}},\ }\href {https://doi.org/10.1063/5.0046032} {\bibfield  {journal}
  {\bibinfo  {journal} {Appl. Phys. Lett.}\ }\textbf {\bibinfo {volume}
  {118}},\ \bibinfo {pages} {202405} (\bibinfo {year} {2021})}\BibitemShut
  {NoStop}%
\bibitem [{\citenamefont {Williame}\ and\ \citenamefont
  {Kim}(2021)}]{Williame2021}%
  \BibitemOpen
  \bibfield  {author} {\bibinfo {author} {\bibfnamefont {J.}~\bibnamefont
  {Williame}}\ and\ \bibinfo {author} {\bibfnamefont {J.-V.}\ \bibnamefont
  {Kim}},\ }\bibfield  {title} {\bibinfo {title} {{A magnetic domain wall
  Mackey–Glass oscillator}},\ }\href {https://doi.org/10.1063/5.0048899}
  {\bibfield  {journal} {\bibinfo  {journal} {Appl. Phys. Lett.}\ }\textbf
  {\bibinfo {volume} {118}},\ \bibinfo {pages} {152404} (\bibinfo {year}
  {2021})}\BibitemShut {NoStop}%
\bibitem [{\citenamefont {Ababei}\ \emph {et~al.}(2021)\citenamefont {Ababei},
  \citenamefont {Ellis}, \citenamefont {Vidamour}, \citenamefont {Devadasan},
  \citenamefont {Allwood}, \citenamefont {Vasilaki},\ and\ \citenamefont
  {Hayward}}]{Ababei2021}%
  \BibitemOpen
  \bibfield  {author} {\bibinfo {author} {\bibfnamefont {R.~V.}\ \bibnamefont
  {Ababei}}, \bibinfo {author} {\bibfnamefont {M.~O.~A.}\ \bibnamefont
  {Ellis}}, \bibinfo {author} {\bibfnamefont {I.~T.}\ \bibnamefont {Vidamour}},
  \bibinfo {author} {\bibfnamefont {D.~S.}\ \bibnamefont {Devadasan}}, \bibinfo
  {author} {\bibfnamefont {D.~A.}\ \bibnamefont {Allwood}}, \bibinfo {author}
  {\bibfnamefont {E.}~\bibnamefont {Vasilaki}},\ and\ \bibinfo {author}
  {\bibfnamefont {T.~J.}\ \bibnamefont {Hayward}},\ }\bibfield  {title}
  {\bibinfo {title} {{Neuromorphic computation with a single magnetic domain
  wall}},\ }\href {https://doi.org/10.1038/s41598-021-94975-y} {\bibfield
  {journal} {\bibinfo  {journal} {Sci. Rep.}\ }\textbf {\bibinfo {volume}
  {11}},\ \bibinfo {pages} {15587} (\bibinfo {year} {2021})}\BibitemShut
  {NoStop}%
\bibitem [{\citenamefont {Siddiqui}\ \emph {et~al.}(2020)\citenamefont
  {Siddiqui}, \citenamefont {Dutta}, \citenamefont {Tang}, \citenamefont {Liu},
  \citenamefont {Ross},\ and\ \citenamefont {Baldo}}]{Siddiqui2020}%
  \BibitemOpen
  \bibfield  {author} {\bibinfo {author} {\bibfnamefont {S.~A.}\ \bibnamefont
  {Siddiqui}}, \bibinfo {author} {\bibfnamefont {S.}~\bibnamefont {Dutta}},
  \bibinfo {author} {\bibfnamefont {A.}~\bibnamefont {Tang}}, \bibinfo {author}
  {\bibfnamefont {L.}~\bibnamefont {Liu}}, \bibinfo {author} {\bibfnamefont
  {C.~A.}\ \bibnamefont {Ross}},\ and\ \bibinfo {author} {\bibfnamefont
  {M.~A.}\ \bibnamefont {Baldo}},\ }\bibfield  {title} {\bibinfo {title}
  {{Magnetic Domain Wall Based Synaptic and Activation Function Generator for
  Neuromorphic Accelerators}},\ }\href
  {https://doi.org/10.1021/acs.nanolett.9b04200} {\bibfield  {journal}
  {\bibinfo  {journal} {Nano Lett.}\ }\textbf {\bibinfo {volume} {20}},\
  \bibinfo {pages} {1033} (\bibinfo {year} {2020})}\BibitemShut {NoStop}%
\bibitem [{\citenamefont {Hassan}\ \emph {et~al.}(2018)\citenamefont {Hassan},
  \citenamefont {Hu}, \citenamefont {Jiang-Wei}, \citenamefont {Brigner},
  \citenamefont {Akinola}, \citenamefont {Garcia-Sanchez}, \citenamefont
  {Pasquale}, \citenamefont {Bennett}, \citenamefont {Incorvia},\ and\
  \citenamefont {Friedman}}]{Hassan2018}%
  \BibitemOpen
  \bibfield  {author} {\bibinfo {author} {\bibfnamefont {N.}~\bibnamefont
  {Hassan}}, \bibinfo {author} {\bibfnamefont {X.}~\bibnamefont {Hu}}, \bibinfo
  {author} {\bibfnamefont {L.}~\bibnamefont {Jiang-Wei}}, \bibinfo {author}
  {\bibfnamefont {W.~H.}\ \bibnamefont {Brigner}}, \bibinfo {author}
  {\bibfnamefont {O.~G.}\ \bibnamefont {Akinola}}, \bibinfo {author}
  {\bibfnamefont {F.}~\bibnamefont {Garcia-Sanchez}}, \bibinfo {author}
  {\bibfnamefont {M.}~\bibnamefont {Pasquale}}, \bibinfo {author}
  {\bibfnamefont {C.~H.}\ \bibnamefont {Bennett}}, \bibinfo {author}
  {\bibfnamefont {J.~A.~C.}\ \bibnamefont {Incorvia}},\ and\ \bibinfo {author}
  {\bibfnamefont {J.~S.}\ \bibnamefont {Friedman}},\ }\bibfield  {title}
  {\bibinfo {title} {{Magnetic domain wall neuron with lateral inhibition}},\
  }\href {https://doi.org/10.1063/1.5042452} {\bibfield  {journal} {\bibinfo
  {journal} {J. Appl. Phys.}\ }\textbf {\bibinfo {volume} {124}},\ \bibinfo
  {pages} {152127} (\bibinfo {year} {2018})}\BibitemShut {NoStop}%
\bibitem [{\citenamefont {Shibata}\ \emph {et~al.}(2011)\citenamefont
  {Shibata}, \citenamefont {Tatara},\ and\ \citenamefont
  {Kohno}}]{Shibata2011}%
  \BibitemOpen
  \bibfield  {author} {\bibinfo {author} {\bibfnamefont {J.}~\bibnamefont
  {Shibata}}, \bibinfo {author} {\bibfnamefont {G.}~\bibnamefont {Tatara}},\
  and\ \bibinfo {author} {\bibfnamefont {H.}~\bibnamefont {Kohno}},\ }\bibfield
   {title} {\bibinfo {title} {{A brief review of field- and current-driven
  domain-wall motion}},\ }\href
  {https://doi.org/10.1088/0022-3727/44/38/384004} {\bibfield  {journal}
  {\bibinfo  {journal} {J. Phys. D: Appl. Phys.}\ }\textbf {\bibinfo {volume}
  {44}},\ \bibinfo {pages} {384004} (\bibinfo {year} {2011})}\BibitemShut
  {NoStop}%
\bibitem [{\citenamefont {Beach}\ \emph {et~al.}(2005)\citenamefont {Beach},
  \citenamefont {Nistor}, \citenamefont {Knutson}, \citenamefont {Tsoi},\ and\
  \citenamefont {Erskine}}]{Beach2005}%
  \BibitemOpen
  \bibfield  {author} {\bibinfo {author} {\bibfnamefont {G.~S.~D.}\
  \bibnamefont {Beach}}, \bibinfo {author} {\bibfnamefont {C.}~\bibnamefont
  {Nistor}}, \bibinfo {author} {\bibfnamefont {C.}~\bibnamefont {Knutson}},
  \bibinfo {author} {\bibfnamefont {M.}~\bibnamefont {Tsoi}},\ and\ \bibinfo
  {author} {\bibfnamefont {J.~L.}\ \bibnamefont {Erskine}},\ }\bibfield
  {title} {\bibinfo {title} {{Dynamics of field-driven domain-wall propagation
  in ferromagnetic nanowires}},\ }\href {https://doi.org/10.1038/nmat1477}
  {\bibfield  {journal} {\bibinfo  {journal} {Nat. Mater.}\ }\textbf {\bibinfo
  {volume} {4}},\ \bibinfo {pages} {741} (\bibinfo {year} {2005})}\BibitemShut
  {NoStop}%
\bibitem [{\citenamefont {Thiaville}\ \emph {et~al.}(2005)\citenamefont
  {Thiaville}, \citenamefont {Nakatani}, \citenamefont {Miltat},\ and\
  \citenamefont {Suzuki}}]{Thiaville2005a}%
  \BibitemOpen
  \bibfield  {author} {\bibinfo {author} {\bibfnamefont {A.}~\bibnamefont
  {Thiaville}}, \bibinfo {author} {\bibfnamefont {Y.}~\bibnamefont {Nakatani}},
  \bibinfo {author} {\bibfnamefont {J.}~\bibnamefont {Miltat}},\ and\ \bibinfo
  {author} {\bibfnamefont {Y.}~\bibnamefont {Suzuki}},\ }\bibfield  {title}
  {\bibinfo {title} {{Micromagnetic understanding of current-driven domain wall
  motion in patterned nanowires}},\ }\href
  {https://doi.org/10.1209/epl/i2004-10452-6} {\bibfield  {journal} {\bibinfo
  {journal} {Europhys. Lett.}\ }\textbf {\bibinfo {volume} {69}},\ \bibinfo
  {pages} {990} (\bibinfo {year} {2005})}\BibitemShut {NoStop}%
\bibitem [{\citenamefont {Manchon}\ \emph {et~al.}(2019)\citenamefont
  {Manchon}, \citenamefont {{\v{Z}}elezn{\'{y}}}, \citenamefont {Miron},
  \citenamefont {Jungwirth}, \citenamefont {Sinova}, \citenamefont {Thiaville},
  \citenamefont {Garello},\ and\ \citenamefont {Gambardella}}]{Manchon2019}%
  \BibitemOpen
  \bibfield  {author} {\bibinfo {author} {\bibfnamefont {A.}~\bibnamefont
  {Manchon}}, \bibinfo {author} {\bibfnamefont {J.}~\bibnamefont
  {{\v{Z}}elezn{\'{y}}}}, \bibinfo {author} {\bibfnamefont {I.~M.}\
  \bibnamefont {Miron}}, \bibinfo {author} {\bibfnamefont {T.}~\bibnamefont
  {Jungwirth}}, \bibinfo {author} {\bibfnamefont {J.}~\bibnamefont {Sinova}},
  \bibinfo {author} {\bibfnamefont {A.}~\bibnamefont {Thiaville}}, \bibinfo
  {author} {\bibfnamefont {K.}~\bibnamefont {Garello}},\ and\ \bibinfo {author}
  {\bibfnamefont {P.}~\bibnamefont {Gambardella}},\ }\bibfield  {title}
  {\bibinfo {title} {{Current-induced spin-orbit torques in ferromagnetic and
  antiferromagnetic systems}},\ }\href
  {https://doi.org/10.1103/RevModPhys.91.035004} {\bibfield  {journal}
  {\bibinfo  {journal} {Rev. Mod. Phys.}\ }\textbf {\bibinfo {volume} {91}},\
  \bibinfo {pages} {035004} (\bibinfo {year} {2019})}\BibitemShut {NoStop}%
\bibitem [{\citenamefont {Khvalkovskiy}\ \emph {et~al.}(2013)\citenamefont
  {Khvalkovskiy}, \citenamefont {Cros}, \citenamefont {Apalkov}, \citenamefont
  {Nikitin}, \citenamefont {Krounbi}, \citenamefont {Zvezdin}, \citenamefont
  {Anane}, \citenamefont {Grollier},\ and\ \citenamefont
  {Fert}}]{Khvalkovskiy2013}%
  \BibitemOpen
  \bibfield  {author} {\bibinfo {author} {\bibfnamefont {A.~V.}\ \bibnamefont
  {Khvalkovskiy}}, \bibinfo {author} {\bibfnamefont {V.}~\bibnamefont {Cros}},
  \bibinfo {author} {\bibfnamefont {D.}~\bibnamefont {Apalkov}}, \bibinfo
  {author} {\bibfnamefont {V.}~\bibnamefont {Nikitin}}, \bibinfo {author}
  {\bibfnamefont {M.}~\bibnamefont {Krounbi}}, \bibinfo {author} {\bibfnamefont
  {K.~A.}\ \bibnamefont {Zvezdin}}, \bibinfo {author} {\bibfnamefont
  {A.}~\bibnamefont {Anane}}, \bibinfo {author} {\bibfnamefont
  {J.}~\bibnamefont {Grollier}},\ and\ \bibinfo {author} {\bibfnamefont
  {A.}~\bibnamefont {Fert}},\ }\bibfield  {title} {\bibinfo {title} {Matching
  domain-wall configuration and spin-orbit torques for efficient domain-wall
  motion},\ }\href {https://doi.org/10.1103/PhysRevB.87.020402} {\bibfield
  {journal} {\bibinfo  {journal} {Phys. Rev. B}\ }\textbf {\bibinfo {volume}
  {87}},\ \bibinfo {pages} {020402} (\bibinfo {year} {2013})}\BibitemShut
  {NoStop}%
\bibitem [{\citenamefont {Ryu}\ \emph {et~al.}(2013)\citenamefont {Ryu},
  \citenamefont {Thomas}, \citenamefont {Yang},\ and\ \citenamefont
  {Parkin}}]{Ryu2013}%
  \BibitemOpen
  \bibfield  {author} {\bibinfo {author} {\bibfnamefont {K.-S.}\ \bibnamefont
  {Ryu}}, \bibinfo {author} {\bibfnamefont {L.}~\bibnamefont {Thomas}},
  \bibinfo {author} {\bibfnamefont {S.-H.}\ \bibnamefont {Yang}},\ and\
  \bibinfo {author} {\bibfnamefont {S.}~\bibnamefont {Parkin}},\ }\bibfield
  {title} {\bibinfo {title} {{Chiral spin torque at magnetic domain walls}},\
  }\href {https://doi.org/10.1038/nnano.2013.102} {\bibfield  {journal}
  {\bibinfo  {journal} {Nat. Nanotech.}\ }\textbf {\bibinfo {volume} {8}},\
  \bibinfo {pages} {527} (\bibinfo {year} {2013})}\BibitemShut {NoStop}%
\bibitem [{\citenamefont {Thiaville}\ \emph {et~al.}(2012)\citenamefont
  {Thiaville}, \citenamefont {Rohart}, \citenamefont {Émilie Ju\'{e}},
  \citenamefont {Cros},\ and\ \citenamefont {Fert}}]{Thiaville2012}%
  \BibitemOpen
  \bibfield  {author} {\bibinfo {author} {\bibfnamefont {A.}~\bibnamefont
  {Thiaville}}, \bibinfo {author} {\bibfnamefont {S.}~\bibnamefont {Rohart}},
  \bibinfo {author} {\bibnamefont {Émilie Ju\'{e}}}, \bibinfo {author}
  {\bibfnamefont {V.}~\bibnamefont {Cros}},\ and\ \bibinfo {author}
  {\bibfnamefont {A.}~\bibnamefont {Fert}},\ }\bibfield  {title} {\bibinfo
  {title} {Dynamics of {D}zyaloshinskii domain walls in ultrathin magnetic
  films},\ }\href {https://doi.org/10.1209/0295-5075/100/57002} {\bibfield
  {journal} {\bibinfo  {journal} {Europhys. Lett.}\ }\textbf {\bibinfo {volume}
  {100}},\ \bibinfo {pages} {57002} (\bibinfo {year} {2012})}\BibitemShut
  {NoStop}%
\bibitem [{\citenamefont {Chen}\ \emph {et~al.}(2013)\citenamefont {Chen},
  \citenamefont {Ma}, \citenamefont {N’Diaye}, \citenamefont {Kwon},
  \citenamefont {Won}, \citenamefont {Wu},\ and\ \citenamefont
  {Schmid}}]{Chen2013}%
  \BibitemOpen
  \bibfield  {author} {\bibinfo {author} {\bibfnamefont {G.}~\bibnamefont
  {Chen}}, \bibinfo {author} {\bibfnamefont {T.}~\bibnamefont {Ma}}, \bibinfo
  {author} {\bibfnamefont {A.~T.}\ \bibnamefont {N’Diaye}}, \bibinfo {author}
  {\bibfnamefont {H.}~\bibnamefont {Kwon}}, \bibinfo {author} {\bibfnamefont
  {C.}~\bibnamefont {Won}}, \bibinfo {author} {\bibfnamefont {Y.}~\bibnamefont
  {Wu}},\ and\ \bibinfo {author} {\bibfnamefont {A.~K.}\ \bibnamefont
  {Schmid}},\ }\bibfield  {title} {\bibinfo {title} {Tailoring the chirality of
  magnetic domain walls by interface engineering},\ }\href@noop {} {\bibfield
  {journal} {\bibinfo  {journal} {Nat. Commun.}\ }\textbf {\bibinfo {volume}
  {4}},\ \bibinfo {pages} {2671} (\bibinfo {year} {2013})}\BibitemShut
  {NoStop}%
\bibitem [{\citenamefont {Torrejon}\ \emph {et~al.}(2014)\citenamefont
  {Torrejon}, \citenamefont {Kim}, \citenamefont {Sinha}, \citenamefont
  {Mitani}, \citenamefont {Hayashi}, \citenamefont {Yamanouchi},\ and\
  \citenamefont {Ohno}}]{Torrejon2014}%
  \BibitemOpen
  \bibfield  {author} {\bibinfo {author} {\bibfnamefont {J.}~\bibnamefont
  {Torrejon}}, \bibinfo {author} {\bibfnamefont {J.}~\bibnamefont {Kim}},
  \bibinfo {author} {\bibfnamefont {J.}~\bibnamefont {Sinha}}, \bibinfo
  {author} {\bibfnamefont {S.}~\bibnamefont {Mitani}}, \bibinfo {author}
  {\bibfnamefont {M.}~\bibnamefont {Hayashi}}, \bibinfo {author} {\bibfnamefont
  {M.}~\bibnamefont {Yamanouchi}},\ and\ \bibinfo {author} {\bibfnamefont
  {H.}~\bibnamefont {Ohno}},\ }\bibfield  {title} {\bibinfo {title} {Interface
  control of the magnetic chirality in {CoFeB/MgO} heterostructures with
  heavy-metal underlayers},\ }\href@noop {} {\bibfield  {journal} {\bibinfo
  {journal} {Nat. Commun.}\ }\textbf {\bibinfo {volume} {5}},\ \bibinfo {pages}
  {4655} (\bibinfo {year} {2014})}\BibitemShut {NoStop}%
\bibitem [{\citenamefont {Hrabec}\ \emph {et~al.}(2014)\citenamefont {Hrabec},
  \citenamefont {Porter}, \citenamefont {Wells}, \citenamefont {Benitez},
  \citenamefont {Burnell}, \citenamefont {McVitie}, \citenamefont {McGrouther},
  \citenamefont {Moore},\ and\ \citenamefont {Marrows}}]{Hrabec2014}%
  \BibitemOpen
  \bibfield  {author} {\bibinfo {author} {\bibfnamefont {A.}~\bibnamefont
  {Hrabec}}, \bibinfo {author} {\bibfnamefont {N.~A.}\ \bibnamefont {Porter}},
  \bibinfo {author} {\bibfnamefont {A.}~\bibnamefont {Wells}}, \bibinfo
  {author} {\bibfnamefont {M.~J.}\ \bibnamefont {Benitez}}, \bibinfo {author}
  {\bibfnamefont {G.}~\bibnamefont {Burnell}}, \bibinfo {author} {\bibfnamefont
  {S.}~\bibnamefont {McVitie}}, \bibinfo {author} {\bibfnamefont
  {D.}~\bibnamefont {McGrouther}}, \bibinfo {author} {\bibfnamefont {T.~A.}\
  \bibnamefont {Moore}},\ and\ \bibinfo {author} {\bibfnamefont {C.~H.}\
  \bibnamefont {Marrows}},\ }\bibfield  {title} {\bibinfo {title} {Measuring
  and tailoring the {D}zyaloshinskii-{M}oriya interaction in perpendicularly
  magnetized thin films},\ }\href {https://doi.org/10.1103/PhysRevB.90.020402}
  {\bibfield  {journal} {\bibinfo  {journal} {Phys. Rev. B}\ }\textbf {\bibinfo
  {volume} {90}},\ \bibinfo {pages} {020402} (\bibinfo {year}
  {2014})}\BibitemShut {NoStop}%
\bibitem [{\citenamefont {Emori}\ \emph {et~al.}(2013)\citenamefont {Emori},
  \citenamefont {Bauer}, \citenamefont {Ahn}, \citenamefont {Martinez},\ and\
  \citenamefont {Beach}}]{Emori2013}%
  \BibitemOpen
  \bibfield  {author} {\bibinfo {author} {\bibfnamefont {S.}~\bibnamefont
  {Emori}}, \bibinfo {author} {\bibfnamefont {U.}~\bibnamefont {Bauer}},
  \bibinfo {author} {\bibfnamefont {S.-M.}\ \bibnamefont {Ahn}}, \bibinfo
  {author} {\bibfnamefont {E.}~\bibnamefont {Martinez}},\ and\ \bibinfo
  {author} {\bibfnamefont {G.~S.~D.}\ \bibnamefont {Beach}},\ }\bibfield
  {title} {\bibinfo {title} {{Current-driven dynamics of chiral ferromagnetic
  domain walls}},\ }\href {https://doi.org/10.1038/nmat3675} {\bibfield
  {journal} {\bibinfo  {journal} {Nat. Mater.}\ }\textbf {\bibinfo {volume}
  {12}},\ \bibinfo {pages} {611} (\bibinfo {year} {2013})}\BibitemShut
  {NoStop}%
\bibitem [{\citenamefont {Franke}\ \emph {et~al.}(2021)\citenamefont {Franke},
  \citenamefont {Ophus}, \citenamefont {Schmid},\ and\ \citenamefont
  {Marrows}}]{Franke2021}%
  \BibitemOpen
  \bibfield  {author} {\bibinfo {author} {\bibfnamefont {K.~J.~A.}\
  \bibnamefont {Franke}}, \bibinfo {author} {\bibfnamefont {C.}~\bibnamefont
  {Ophus}}, \bibinfo {author} {\bibfnamefont {A.~K.}\ \bibnamefont {Schmid}},\
  and\ \bibinfo {author} {\bibfnamefont {C.~H.}\ \bibnamefont {Marrows}},\
  }\bibfield  {title} {\bibinfo {title} {Switching between magnetic {B}loch and
  {N}\'{e}el domain walls with anisotropy modulations},\ }\href
  {https://doi.org/10.1103/PhysRevLett.127.127203} {\bibfield  {journal}
  {\bibinfo  {journal} {Phys. Rev. Lett.}\ }\textbf {\bibinfo {volume} {127}},\
  \bibinfo {pages} {127203} (\bibinfo {year} {2021})}\BibitemShut {NoStop}%
\bibitem [{\citenamefont {Wu}\ \emph {et~al.}(2015)\citenamefont {Wu},
  \citenamefont {Huang}, \citenamefont {Cui}, \citenamefont {Shen},
  \citenamefont {Ma}, \citenamefont {Xie}, \citenamefont {Du}, \citenamefont
  {Gao},\ and\ \citenamefont {Li}}]{Wu2015}%
  \BibitemOpen
  \bibfield  {author} {\bibinfo {author} {\bibfnamefont {C.}~\bibnamefont
  {Wu}}, \bibinfo {author} {\bibfnamefont {Y.}~\bibnamefont {Huang}}, \bibinfo
  {author} {\bibfnamefont {Y.}~\bibnamefont {Cui}}, \bibinfo {author}
  {\bibfnamefont {Z.}~\bibnamefont {Shen}}, \bibinfo {author} {\bibfnamefont
  {Y.}~\bibnamefont {Ma}}, \bibinfo {author} {\bibfnamefont {S.}~\bibnamefont
  {Xie}}, \bibinfo {author} {\bibfnamefont {H.}~\bibnamefont {Du}}, \bibinfo
  {author} {\bibfnamefont {X.}~\bibnamefont {Gao}},\ and\ \bibinfo {author}
  {\bibfnamefont {S.}~\bibnamefont {Li}},\ }\bibfield  {title} {\bibinfo
  {title} {{Film Thickness Gradient-Induced Magnetic Anisotropy and
  Ferromagnetic Resonance in {Fe}$_{56}${Co}$_{24}${B}$_{2}$Amorphous Films
  Prepared by Pulse Laser Deposition}},\ }\href
  {https://doi.org/10.1109/TMAG.2015.2437912} {\bibfield  {journal} {\bibinfo
  {journal} {IEEE Trans. Magn.}\ }\textbf {\bibinfo {volume} {51}},\ \bibinfo
  {pages} {1} (\bibinfo {year} {2015})}\BibitemShut {NoStop}%
\bibitem [{\citenamefont {Chen}\ \emph {et~al.}(2015)\citenamefont {Chen},
  \citenamefont {N'Diaye}, \citenamefont {Kang}, \citenamefont {Kwon},
  \citenamefont {Won}, \citenamefont {Wu}, \citenamefont {Qiu},\ and\
  \citenamefont {Schmid}}]{Chen2015}%
  \BibitemOpen
  \bibfield  {author} {\bibinfo {author} {\bibfnamefont {G.}~\bibnamefont
  {Chen}}, \bibinfo {author} {\bibfnamefont {A.~T.}\ \bibnamefont {N'Diaye}},
  \bibinfo {author} {\bibfnamefont {S.~P.}\ \bibnamefont {Kang}}, \bibinfo
  {author} {\bibfnamefont {H.~Y.}\ \bibnamefont {Kwon}}, \bibinfo {author}
  {\bibfnamefont {C.}~\bibnamefont {Won}}, \bibinfo {author} {\bibfnamefont
  {Y.}~\bibnamefont {Wu}}, \bibinfo {author} {\bibfnamefont {Z.~Q.}\
  \bibnamefont {Qiu}},\ and\ \bibinfo {author} {\bibfnamefont {A.~K.}\
  \bibnamefont {Schmid}},\ }\bibfield  {title} {\bibinfo {title} {{Unlocking
  Bloch-type chirality in ultrathin magnets through uniaxial strain}},\ }\href
  {https://doi.org/10.1038/ncomms7598} {\bibfield  {journal} {\bibinfo
  {journal} {Nat. Commun.}\ }\textbf {\bibinfo {volume} {6}},\ \bibinfo {pages}
  {6598} (\bibinfo {year} {2015})}\BibitemShut {NoStop}%
\bibitem [{\citenamefont {DeJong}\ and\ \citenamefont
  {Livesey}(2015)}]{Dejong2015a}%
  \BibitemOpen
  \bibfield  {author} {\bibinfo {author} {\bibfnamefont {M.~D.}\ \bibnamefont
  {DeJong}}\ and\ \bibinfo {author} {\bibfnamefont {K.~L.}\ \bibnamefont
  {Livesey}},\ }\bibfield  {title} {\bibinfo {title} {{Analytic theory for the
  switch from Bloch to N{\'{e}}el domain wall in nanowires with perpendicular
  anisotropy}},\ }\href {https://doi.org/10.1103/PhysRevB.92.214420} {\bibfield
   {journal} {\bibinfo  {journal} {Phys. Rev. B}\ }\textbf {\bibinfo {volume}
  {92}},\ \bibinfo {pages} {214420} (\bibinfo {year} {2015})}\BibitemShut
  {NoStop}%
\bibitem [{\citenamefont {Belyaev}\ \emph {et~al.}(2020)\citenamefont
  {Belyaev}, \citenamefont {Izotov}, \citenamefont {Solovev},\ and\
  \citenamefont {Boev}}]{Belyaev2020}%
  \BibitemOpen
  \bibfield  {author} {\bibinfo {author} {\bibfnamefont {B.~A.}\ \bibnamefont
  {Belyaev}}, \bibinfo {author} {\bibfnamefont {A.~V.}\ \bibnamefont {Izotov}},
  \bibinfo {author} {\bibfnamefont {P.~N.}\ \bibnamefont {Solovev}},\ and\
  \bibinfo {author} {\bibfnamefont {N.~M.}\ \bibnamefont {Boev}},\ }\bibfield
  {title} {\bibinfo {title} {{Strain‐Gradient‐Induced Unidirectional
  Magnetic Anisotropy in Nanocrystalline Thin Permalloy Films}},\ }\href
  {https://doi.org/10.1002/pssr.201900467} {\bibfield  {journal} {\bibinfo
  {journal} {Phys. Status Solidi RRL}\ }\textbf {\bibinfo {volume} {14}},\
  \bibinfo {pages} {1900467} (\bibinfo {year} {2020})}\BibitemShut {NoStop}%
\bibitem [{\citenamefont {Fattouhi}\ \emph {et~al.}(2022)\citenamefont
  {Fattouhi}, \citenamefont {Garcia-Sanchez}, \citenamefont {Yanes},
  \citenamefont {Raposo}, \citenamefont {Martinez},\ and\ \citenamefont
  {Lopez-Diaz}}]{Fattouhi2022}%
  \BibitemOpen
  \bibfield  {author} {\bibinfo {author} {\bibfnamefont {M.}~\bibnamefont
  {Fattouhi}}, \bibinfo {author} {\bibfnamefont {F.}~\bibnamefont
  {Garcia-Sanchez}}, \bibinfo {author} {\bibfnamefont {R.}~\bibnamefont
  {Yanes}}, \bibinfo {author} {\bibfnamefont {V.}~\bibnamefont {Raposo}},
  \bibinfo {author} {\bibfnamefont {E.}~\bibnamefont {Martinez}},\ and\
  \bibinfo {author} {\bibfnamefont {L.}~\bibnamefont {Lopez-Diaz}},\ }\bibfield
   {title} {\bibinfo {title} {Absence of walker breakdown in the dynamics of
  chiral néel domain walls driven by in-plane strain gradients},\ }\href
  {https://doi.org/10.1103/PhysRevApplied.18.044023} {\bibfield  {journal}
  {\bibinfo  {journal} {Physical Review Applied}\ }\textbf {\bibinfo {volume}
  {18}},\ \bibinfo {pages} {044023} (\bibinfo {year} {2022})}\BibitemShut
  {NoStop}%
\bibitem [{\citenamefont {Omari}\ \emph {et~al.}(2019)\citenamefont {Omari},
  \citenamefont {Broomhall}, \citenamefont {Dawidek}, \citenamefont {Allwood},
  \citenamefont {Bradley}, \citenamefont {Wood}, \citenamefont {Fry},
  \citenamefont {Rosamond}, \citenamefont {Linfield}, \citenamefont {Im},
  \citenamefont {Fischer},\ and\ \citenamefont {Hayward}}]{Omari2019}%
  \BibitemOpen
  \bibfield  {author} {\bibinfo {author} {\bibfnamefont {K.~A.}\ \bibnamefont
  {Omari}}, \bibinfo {author} {\bibfnamefont {T.~J.}\ \bibnamefont
  {Broomhall}}, \bibinfo {author} {\bibfnamefont {R.~W.}\ \bibnamefont
  {Dawidek}}, \bibinfo {author} {\bibfnamefont {D.~A.}\ \bibnamefont
  {Allwood}}, \bibinfo {author} {\bibfnamefont {R.~C.}\ \bibnamefont
  {Bradley}}, \bibinfo {author} {\bibfnamefont {J.~M.}\ \bibnamefont {Wood}},
  \bibinfo {author} {\bibfnamefont {P.~W.}\ \bibnamefont {Fry}}, \bibinfo
  {author} {\bibfnamefont {M.~C.}\ \bibnamefont {Rosamond}}, \bibinfo {author}
  {\bibfnamefont {E.~H.}\ \bibnamefont {Linfield}}, \bibinfo {author}
  {\bibfnamefont {M.~Y.}\ \bibnamefont {Im}}, \bibinfo {author} {\bibfnamefont
  {P.~J.}\ \bibnamefont {Fischer}},\ and\ \bibinfo {author} {\bibfnamefont
  {T.~J.}\ \bibnamefont {Hayward}},\ }\bibfield  {title} {\bibinfo {title}
  {Toward chirality-encoded domain wall logic},\ }\href
  {https://doi.org/10.1002/adfm.201807282} {\bibfield  {journal} {\bibinfo
  {journal} {Advanced Functional Materials}\ }\textbf {\bibinfo {volume}
  {29}},\ \bibinfo {pages} {1807282} (\bibinfo {year} {2019})}\BibitemShut
  {NoStop}%
\bibitem [{\citenamefont {Skaugen}\ \emph {et~al.}(2019)\citenamefont
  {Skaugen}, \citenamefont {Murray},\ and\ \citenamefont
  {Laurson}}]{Skaugen2019}%
  \BibitemOpen
  \bibfield  {author} {\bibinfo {author} {\bibfnamefont {A.}~\bibnamefont
  {Skaugen}}, \bibinfo {author} {\bibfnamefont {P.}~\bibnamefont {Murray}},\
  and\ \bibinfo {author} {\bibfnamefont {L.}~\bibnamefont {Laurson}},\
  }\bibfield  {title} {\bibinfo {title} {{Analytical computation of the
  demagnetizing energy of thin-film domain walls}},\ }\href
  {https://doi.org/10.1103/PhysRevB.100.094440} {\bibfield  {journal} {\bibinfo
   {journal} {Phys. Rev. B}\ }\textbf {\bibinfo {volume} {100}},\ \bibinfo
  {pages} {094440} (\bibinfo {year} {2019})}\BibitemShut {NoStop}%
\bibitem [{\citenamefont {Boehm}\ \emph {et~al.}(2017)\citenamefont {Boehm},
  \citenamefont {Bisig}, \citenamefont {Bischof}, \citenamefont {Stefanou},
  \citenamefont {Hickey},\ and\ \citenamefont {Allenspach}}]{Boehm2017}%
  \BibitemOpen
  \bibfield  {author} {\bibinfo {author} {\bibfnamefont {B.}~\bibnamefont
  {Boehm}}, \bibinfo {author} {\bibfnamefont {A.}~\bibnamefont {Bisig}},
  \bibinfo {author} {\bibfnamefont {A.}~\bibnamefont {Bischof}}, \bibinfo
  {author} {\bibfnamefont {G.}~\bibnamefont {Stefanou}}, \bibinfo {author}
  {\bibfnamefont {B.~J.}\ \bibnamefont {Hickey}},\ and\ \bibinfo {author}
  {\bibfnamefont {R.}~\bibnamefont {Allenspach}},\ }\bibfield  {title}
  {\bibinfo {title} {{Achiral tilted domain walls in perpendicularly magnetized
  nanowires}},\ }\href {https://doi.org/10.1103/PhysRevB.95.180406} {\bibfield
  {journal} {\bibinfo  {journal} {Phys. Rev. B}\ }\textbf {\bibinfo {volume}
  {95}},\ \bibinfo {pages} {1} (\bibinfo {year} {2017})}\BibitemShut {NoStop}%
\bibitem [{\citenamefont {Avci}\ \emph {et~al.}(2019)\citenamefont {Avci},
  \citenamefont {Beach},\ and\ \citenamefont {Gambardella}}]{Avci2019b}%
  \BibitemOpen
  \bibfield  {author} {\bibinfo {author} {\bibfnamefont {C.~O.}\ \bibnamefont
  {Avci}}, \bibinfo {author} {\bibfnamefont {G.~S.~D.}\ \bibnamefont {Beach}},\
  and\ \bibinfo {author} {\bibfnamefont {P.}~\bibnamefont {Gambardella}},\
  }\bibfield  {title} {\bibinfo {title} {{Effects of transition metal spacers
  on spin-orbit torques, spin Hall magnetoresistance, and magnetic anisotropy
  of Pt/Co bilayers}},\ }\href {https://doi.org/10.1103/PhysRevB.100.235454}
  {\bibfield  {journal} {\bibinfo  {journal} {Phys. Rev. B}\ }\textbf {\bibinfo
  {volume} {100}},\ \bibinfo {pages} {235454} (\bibinfo {year}
  {2019})}\BibitemShut {NoStop}%
\bibitem [{\citenamefont {Nan}\ \emph {et~al.}(2015)\citenamefont {Nan},
  \citenamefont {Emori}, \citenamefont {Boone}, \citenamefont {Wang},
  \citenamefont {Oxholm}, \citenamefont {Jones}, \citenamefont {Howe},
  \citenamefont {Brown},\ and\ \citenamefont {Sun}}]{Nan2015}%
  \BibitemOpen
  \bibfield  {author} {\bibinfo {author} {\bibfnamefont {T.}~\bibnamefont
  {Nan}}, \bibinfo {author} {\bibfnamefont {S.}~\bibnamefont {Emori}}, \bibinfo
  {author} {\bibfnamefont {C.~T.}\ \bibnamefont {Boone}}, \bibinfo {author}
  {\bibfnamefont {X.}~\bibnamefont {Wang}}, \bibinfo {author} {\bibfnamefont
  {T.~M.}\ \bibnamefont {Oxholm}}, \bibinfo {author} {\bibfnamefont {J.~G.}\
  \bibnamefont {Jones}}, \bibinfo {author} {\bibfnamefont {B.~M.}\ \bibnamefont
  {Howe}}, \bibinfo {author} {\bibfnamefont {G.~J.}\ \bibnamefont {Brown}},\
  and\ \bibinfo {author} {\bibfnamefont {N.~X.}\ \bibnamefont {Sun}},\
  }\bibfield  {title} {\bibinfo {title} {{Comparison of spin-orbit torques and
  spin pumping across NiFe/Pt and NiFe/Cu/Pt interfaces}},\ }\href
  {https://doi.org/10.1103/PhysRevB.91.214416} {\bibfield  {journal} {\bibinfo
  {journal} {Phys. Rev. B}\ }\textbf {\bibinfo {volume} {91}},\ \bibinfo
  {pages} {214416} (\bibinfo {year} {2015})}\BibitemShut {NoStop}%
\bibitem [{\citenamefont {Chen}\ \emph {et~al.}(2002)\citenamefont {Chen},
  \citenamefont {Pardo},\ and\ \citenamefont {Sanchez}}]{Chen2002}%
  \BibitemOpen
  \bibfield  {author} {\bibinfo {author} {\bibfnamefont {D.~X.}\ \bibnamefont
  {Chen}}, \bibinfo {author} {\bibfnamefont {E.}~\bibnamefont {Pardo}},\ and\
  \bibinfo {author} {\bibfnamefont {A.}~\bibnamefont {Sanchez}},\ }\bibfield
  {title} {\bibinfo {title} {{Demagnetizing factors of rectangular prisms and
  ellipsoids}},\ }\href {https://doi.org/10.1109/TMAG.2002.1017766} {\bibfield
  {journal} {\bibinfo  {journal} {IEEE Trans. Magn.}\ }\textbf {\bibinfo
  {volume} {38}},\ \bibinfo {pages} {1742} (\bibinfo {year}
  {2002})}\BibitemShut {NoStop}%
\bibitem [{\citenamefont {Mougin}\ \emph {et~al.}(2007)\citenamefont {Mougin},
  \citenamefont {Cormier}, \citenamefont {Adam}, \citenamefont {Metaxas},\ and\
  \citenamefont {Ferr{\'{e}}}}]{Mougin2007}%
  \BibitemOpen
  \bibfield  {author} {\bibinfo {author} {\bibfnamefont {A.}~\bibnamefont
  {Mougin}}, \bibinfo {author} {\bibfnamefont {M.}~\bibnamefont {Cormier}},
  \bibinfo {author} {\bibfnamefont {J.~P.}\ \bibnamefont {Adam}}, \bibinfo
  {author} {\bibfnamefont {P.~J.}\ \bibnamefont {Metaxas}},\ and\ \bibinfo
  {author} {\bibfnamefont {J.}~\bibnamefont {Ferr{\'{e}}}},\ }\bibfield
  {title} {\bibinfo {title} {{Domain wall mobility, stability and Walker
  breakdown in magnetic nanowires}},\ }\href
  {https://doi.org/10.1209/0295-5075/78/57007} {\bibfield  {journal} {\bibinfo
  {journal} {Europhys. Lett.}\ }\textbf {\bibinfo {volume} {78}},\ \bibinfo
  {pages} {57007} (\bibinfo {year} {2007})}\BibitemShut {NoStop}%
\bibitem [{\citenamefont {You}(2006)}]{You2006b}%
  \BibitemOpen
  \bibfield  {author} {\bibinfo {author} {\bibfnamefont {C.-Y.}\ \bibnamefont
  {You}},\ }\bibfield  {title} {\bibinfo {title} {{Confined magnetic stray
  field from a narrow domain wall}},\ }\href
  {https://doi.org/10.1063/1.2266233} {\bibfield  {journal} {\bibinfo
  {journal} {J. Appl. Phys.}\ }\textbf {\bibinfo {volume} {100}},\ \bibinfo
  {pages} {043911} (\bibinfo {year} {2006})}\BibitemShut {NoStop}%
\bibitem [{\citenamefont {Vansteenkiste}\ \emph {et~al.}(2014)\citenamefont
  {Vansteenkiste}, \citenamefont {Leliaert}, \citenamefont {Dvornik},
  \citenamefont {Helsen}, \citenamefont {Garcia-Sanchez},\ and\ \citenamefont
  {{Van Waeyenberge}}}]{Vansteenkiste2014}%
  \BibitemOpen
  \bibfield  {author} {\bibinfo {author} {\bibfnamefont {A.}~\bibnamefont
  {Vansteenkiste}}, \bibinfo {author} {\bibfnamefont {J.}~\bibnamefont
  {Leliaert}}, \bibinfo {author} {\bibfnamefont {M.}~\bibnamefont {Dvornik}},
  \bibinfo {author} {\bibfnamefont {M.}~\bibnamefont {Helsen}}, \bibinfo
  {author} {\bibfnamefont {F.}~\bibnamefont {Garcia-Sanchez}},\ and\ \bibinfo
  {author} {\bibfnamefont {B.}~\bibnamefont {{Van Waeyenberge}}},\ }\bibfield
  {title} {\bibinfo {title} {{The design and verification of MuMax3}},\ }\href
  {https://doi.org/10.1063/1.4899186} {\bibfield  {journal} {\bibinfo
  {journal} {AIP Adv.}\ }\textbf {\bibinfo {volume} {4}},\ \bibinfo {pages}
  {107133} (\bibinfo {year} {2014})}\BibitemShut {NoStop}%
\bibitem [{\citenamefont {Devolder}\ \emph {et~al.}(2016)\citenamefont
  {Devolder}, \citenamefont {Kim}, \citenamefont {Nistor}, \citenamefont
  {Sousa}, \citenamefont {Rodmacq},\ and\ \citenamefont
  {Di{\'{e}}ny}}]{Devolder2016}%
  \BibitemOpen
  \bibfield  {author} {\bibinfo {author} {\bibfnamefont {T.}~\bibnamefont
  {Devolder}}, \bibinfo {author} {\bibfnamefont {J.~V.}\ \bibnamefont {Kim}},
  \bibinfo {author} {\bibfnamefont {L.}~\bibnamefont {Nistor}}, \bibinfo
  {author} {\bibfnamefont {R.}~\bibnamefont {Sousa}}, \bibinfo {author}
  {\bibfnamefont {B.}~\bibnamefont {Rodmacq}},\ and\ \bibinfo {author}
  {\bibfnamefont {B.}~\bibnamefont {Di{\'{e}}ny}},\ }\bibfield  {title}
  {\bibinfo {title} {{Exchange stiffness in ultrathin perpendicularly
  magnetized CoFeB layers determined using the spectroscopy of electrically
  excited spin waves}},\ }\href {https://doi.org/10.1063/1.4967826} {\bibfield
  {journal} {\bibinfo  {journal} {J. Appl. Phys.}\ }\textbf {\bibinfo {volume}
  {120}},\ \bibinfo {pages} {183902} (\bibinfo {year} {2016})}\BibitemShut
  {NoStop}%
\bibitem [{\citenamefont {Schryer}\ and\ \citenamefont
  {Walker}(1974)}]{Schryer1974}%
  \BibitemOpen
  \bibfield  {author} {\bibinfo {author} {\bibfnamefont {N.~L.}\ \bibnamefont
  {Schryer}}\ and\ \bibinfo {author} {\bibfnamefont {L.~R.}\ \bibnamefont
  {Walker}},\ }\bibfield  {title} {\bibinfo {title} {The motion of 180$^\circ$
  domain walls in uniform dc magnetic fields},\ }\href
  {https://doi.org/10.1063/1.1663252} {\bibfield  {journal} {\bibinfo
  {journal} {J. Appl. Phys.}\ }\textbf {\bibinfo {volume} {45}},\ \bibinfo
  {pages} {5406} (\bibinfo {year} {1974})}\BibitemShut {NoStop}%
\bibitem [{\citenamefont {Fukami}\ \emph {et~al.}(2013)\citenamefont {Fukami},
  \citenamefont {Yamanouchi}, \citenamefont {Ikeda},\ and\ \citenamefont
  {Ohno}}]{Fukami2013}%
  \BibitemOpen
  \bibfield  {author} {\bibinfo {author} {\bibfnamefont {S.}~\bibnamefont
  {Fukami}}, \bibinfo {author} {\bibfnamefont {M.}~\bibnamefont {Yamanouchi}},
  \bibinfo {author} {\bibfnamefont {S.}~\bibnamefont {Ikeda}},\ and\ \bibinfo
  {author} {\bibfnamefont {H.}~\bibnamefont {Ohno}},\ }\bibfield  {title}
  {\bibinfo {title} {{Depinning probability of a magnetic domain wall in
  nanowires by spin-polarized currents}},\ }\href
  {https://doi.org/10.1038/ncomms3293} {\bibfield  {journal} {\bibinfo
  {journal} {Nat. Commun.}\ }\textbf {\bibinfo {volume} {4}},\ \bibinfo {pages}
  {2293} (\bibinfo {year} {2013})}\BibitemShut {NoStop}%
\bibitem [{\citenamefont {Hayashi}\ \emph {et~al.}(2008)\citenamefont
  {Hayashi}, \citenamefont {Thomas}, \citenamefont {Rettner}, \citenamefont
  {Moriya},\ and\ \citenamefont {Parkin}}]{Hayashi2008}%
  \BibitemOpen
  \bibfield  {author} {\bibinfo {author} {\bibfnamefont {M.}~\bibnamefont
  {Hayashi}}, \bibinfo {author} {\bibfnamefont {L.}~\bibnamefont {Thomas}},
  \bibinfo {author} {\bibfnamefont {C.}~\bibnamefont {Rettner}}, \bibinfo
  {author} {\bibfnamefont {R.}~\bibnamefont {Moriya}},\ and\ \bibinfo {author}
  {\bibfnamefont {S.~S.~P.}\ \bibnamefont {Parkin}},\ }\bibfield  {title}
  {\bibinfo {title} {{Dynamics of domain wall depinning driven by a combination
  of direct and pulsed currents}},\ }\href {https://doi.org/10.1063/1.2903096}
  {\bibfield  {journal} {\bibinfo  {journal} {Appl. Phys. Lett.}\ }\textbf
  {\bibinfo {volume} {92}},\ \bibinfo {pages} {162503} (\bibinfo {year}
  {2008})}\BibitemShut {NoStop}%
\bibitem [{\citenamefont {Yue}\ \emph {et~al.}(2019)\citenamefont {Yue},
  \citenamefont {Liu}, \citenamefont {Lake},\ and\ \citenamefont
  {Parker}}]{Yue2019}%
  \BibitemOpen
  \bibfield  {author} {\bibinfo {author} {\bibfnamefont {K.}~\bibnamefont
  {Yue}}, \bibinfo {author} {\bibfnamefont {Y.}~\bibnamefont {Liu}}, \bibinfo
  {author} {\bibfnamefont {R.~K.}\ \bibnamefont {Lake}},\ and\ \bibinfo
  {author} {\bibfnamefont {A.~C.}\ \bibnamefont {Parker}},\ }\bibfield  {title}
  {\bibinfo {title} {{A brain-plausible neuromorphic on-the-fly learning system
  implemented with magnetic domain wall analog memristors}},\ }\href
  {https://doi.org/10.1126/sciadv.aau8170} {\bibfield  {journal} {\bibinfo
  {journal} {Sci. Adv.}\ }\textbf {\bibinfo {volume} {5}},\ \bibinfo {pages}
  {1} (\bibinfo {year} {2019})}\BibitemShut {NoStop}%
\bibitem [{\citenamefont {Blachowicz}\ and\ \citenamefont
  {Ehrmann}(2020)}]{Blachowicz2020}%
  \BibitemOpen
  \bibfield  {author} {\bibinfo {author} {\bibfnamefont {T.}~\bibnamefont
  {Blachowicz}}\ and\ \bibinfo {author} {\bibfnamefont {A.}~\bibnamefont
  {Ehrmann}},\ }\bibfield  {title} {\bibinfo {title} {{Magnetic Elements for
  Neuromorphic Computing}},\ }\href {https://doi.org/10.3390/molecules25112550}
  {\bibfield  {journal} {\bibinfo  {journal} {Molecules}\ }\textbf {\bibinfo
  {volume} {25}},\ \bibinfo {pages} {2550} (\bibinfo {year}
  {2020})}\BibitemShut {NoStop}%
\bibitem [{\citenamefont {Markovi{\'{c}}}\ \emph {et~al.}(2020)\citenamefont
  {Markovi{\'{c}}}, \citenamefont {Mizrahi}, \citenamefont {Querlioz},\ and\
  \citenamefont {Grollier}}]{Markovic2020}%
  \BibitemOpen
  \bibfield  {author} {\bibinfo {author} {\bibfnamefont {D.}~\bibnamefont
  {Markovi{\'{c}}}}, \bibinfo {author} {\bibfnamefont {A.}~\bibnamefont
  {Mizrahi}}, \bibinfo {author} {\bibfnamefont {D.}~\bibnamefont {Querlioz}},\
  and\ \bibinfo {author} {\bibfnamefont {J.}~\bibnamefont {Grollier}},\
  }\bibfield  {title} {\bibinfo {title} {{Physics for neuromorphic
  computing}},\ }\href {https://doi.org/10.1038/s42254-020-0208-2} {\bibfield
  {journal} {\bibinfo  {journal} {Nat. Rev. Phys.}\ }\textbf {\bibinfo {volume}
  {2}},\ \bibinfo {pages} {499} (\bibinfo {year} {2020})}\BibitemShut {NoStop}%
\bibitem [{\citenamefont {Toda}(1958)}]{Toda1958}%
  \BibitemOpen
  \bibfield  {author} {\bibinfo {author} {\bibfnamefont {M.}~\bibnamefont
  {Toda}},\ }\bibfield  {title} {\bibinfo {title} {{On the Theory of the
  Brownian Motion}},\ }\href {https://doi.org/10.1143/JPSJ.13.1266} {\bibfield
  {journal} {\bibinfo  {journal} {J. Phys. Soc. Jpn.}\ }\textbf {\bibinfo
  {volume} {13}},\ \bibinfo {pages} {1266} (\bibinfo {year}
  {1958})}\BibitemShut {NoStop}%
\bibitem [{\citenamefont {Diaconis}\ \emph {et~al.}(2007)\citenamefont
  {Diaconis}, \citenamefont {Holmes},\ and\ \citenamefont
  {Montgomery}}]{Diaconis2007}%
  \BibitemOpen
  \bibfield  {author} {\bibinfo {author} {\bibfnamefont {P.}~\bibnamefont
  {Diaconis}}, \bibinfo {author} {\bibfnamefont {S.}~\bibnamefont {Holmes}},\
  and\ \bibinfo {author} {\bibfnamefont {R.}~\bibnamefont {Montgomery}},\
  }\bibfield  {title} {\bibinfo {title} {{Dynamical Bias in the Coin Toss}},\
  }\href {https://doi.org/10.1137/S0036144504446436} {\bibfield  {journal}
  {\bibinfo  {journal} {SIAM Rev.}\ }\textbf {\bibinfo {volume} {49}},\
  \bibinfo {pages} {211} (\bibinfo {year} {2007})}\BibitemShut {NoStop}%
\bibitem [{\citenamefont {Li}\ \emph {et~al.}(2015)\citenamefont {Li},
  \citenamefont {Xue}, \citenamefont {Du}, \citenamefont {Xu}, \citenamefont
  {Li}, \citenamefont {Shi}, \citenamefont {Gao}, \citenamefont {Liu},
  \citenamefont {Nan}, \citenamefont {Hu}, \citenamefont {Sun},\ and\
  \citenamefont {Shao}}]{Li2015}%
  \BibitemOpen
  \bibfield  {author} {\bibinfo {author} {\bibfnamefont {S.}~\bibnamefont
  {Li}}, \bibinfo {author} {\bibfnamefont {Q.}~\bibnamefont {Xue}}, \bibinfo
  {author} {\bibfnamefont {H.}~\bibnamefont {Du}}, \bibinfo {author}
  {\bibfnamefont {J.}~\bibnamefont {Xu}}, \bibinfo {author} {\bibfnamefont
  {Q.}~\bibnamefont {Li}}, \bibinfo {author} {\bibfnamefont {Z.}~\bibnamefont
  {Shi}}, \bibinfo {author} {\bibfnamefont {X.}~\bibnamefont {Gao}}, \bibinfo
  {author} {\bibfnamefont {M.}~\bibnamefont {Liu}}, \bibinfo {author}
  {\bibfnamefont {T.}~\bibnamefont {Nan}}, \bibinfo {author} {\bibfnamefont
  {Z.}~\bibnamefont {Hu}}, \bibinfo {author} {\bibfnamefont {N.~X.}\
  \bibnamefont {Sun}},\ and\ \bibinfo {author} {\bibfnamefont {W.}~\bibnamefont
  {Shao}},\ }\bibfield  {title} {\bibinfo {title} {Large {E}-field tunability
  of magnetic anisotropy and ferromagnetic resonance frequency of co-sputtered
  {Fe}$_{50}${Co}$_{50}$-{B} film},\ }\href {https://doi.org/10.1063/1.4906752}
  {\bibfield  {journal} {\bibinfo  {journal} {J. Appl. Phys.}\ }\textbf
  {\bibinfo {volume} {117}},\ \bibinfo {pages} {17D702} (\bibinfo {year}
  {2015})}\BibitemShut {NoStop}%
\bibitem [{\citenamefont {Lin}\ \emph {et~al.}(2016)\citenamefont {Lin},
  \citenamefont {Vernier}, \citenamefont {Agnus}, \citenamefont {Garcia},
  \citenamefont {Ocker}, \citenamefont {Zhao}, \citenamefont {Fullerton},\ and\
  \citenamefont {Ravelosona}}]{Lin2016}%
  \BibitemOpen
  \bibfield  {author} {\bibinfo {author} {\bibfnamefont {W.}~\bibnamefont
  {Lin}}, \bibinfo {author} {\bibfnamefont {N.}~\bibnamefont {Vernier}},
  \bibinfo {author} {\bibfnamefont {G.}~\bibnamefont {Agnus}}, \bibinfo
  {author} {\bibfnamefont {K.}~\bibnamefont {Garcia}}, \bibinfo {author}
  {\bibfnamefont {B.}~\bibnamefont {Ocker}}, \bibinfo {author} {\bibfnamefont
  {W.}~\bibnamefont {Zhao}}, \bibinfo {author} {\bibfnamefont {E.~E.}\
  \bibnamefont {Fullerton}},\ and\ \bibinfo {author} {\bibfnamefont
  {D.}~\bibnamefont {Ravelosona}},\ }\bibfield  {title} {\bibinfo {title}
  {{Universal domain wall dynamics under electric field in Ta/CoFeB/MgO devices
  with perpendicular anisotropy}},\ }\href
  {https://doi.org/10.1038/ncomms13532} {\bibfield  {journal} {\bibinfo
  {journal} {Nat. Commun.}\ }\textbf {\bibinfo {volume} {7}},\ \bibinfo {pages}
  {13532} (\bibinfo {year} {2016})}\BibitemShut {NoStop}%
\end{thebibliography}%






\end{document}